\documentclass[apsnotitlepage, floatfix, jcp, preprint, ,amsmath,amssymb]{revtex4-1}

\usepackage{graphicx}
\usepackage{graphics}
\usepackage{psfig}

\usepackage{color}

\begin{document}

\newcommand{\Grad}{\nabla_{\bf r_1}}
\newcommand{\Gradi}{\nabla_{\bf r_i}}
\newcommand{\Div}{\nabla_{\bf r_1} \cdot}
\newcommand{\DPT}{\partial_t}
\newcommand{\bu}{{\bf u}}
\newcommand{\DF}{\frac{\delta F[\rho]}{\delta \rho}}
\newcommand{\TT}{\mathcal{T}}
\newcommand{\DD}{\mathcal{D}}
\newcommand{\CC}{{\bf C^{(2)}}}
\newcommand{\normLL}[1]{\parallel #1 \parallel_{L^2}}
\newcommand{\normL}[1]{\parallel #1 \parallel_{L^1}}

\newcommand{\ff}{f^{(1)}}
\newcommand{\ffp}[1]{\ff ( \br_{#1}, \bv_{#1},t ) }
\newcommand{\fff}{f^{(2)}}
\newcommand{\sigh}{\hat{\bsigma}}
\newcommand{\gh}{\hat{\bf g}}
\newcommand{\bv}{{\bf v}}
\newcommand{\bp}{{\bf p}}  
\newcommand{\br}{{\bf r}}
\newcommand{\bg}{{\bf g}}
\newcommand{\bc}{{\bf c}}

\providecommand\bcdot{\boldsymbol{\cdot}}
\providecommand\bnab{\mbox{\boldmath $\nabla$}}
\providecommand\bsigma{\mbox{\boldmath $\sigma$}}
\providecommand\balpha{\mbox{\boldmath $\alpha$}}
\providecommand\bdelta{\mbox{\boldmath $\delta$}}
\providecommand\bepsilon{\mbox{\boldmath $\epsilon$}}
\providecommand\beeta{\mbox{\boldmath $\eta$}}
\providecommand\bIm{\mbox{\boldmath $\Im$}}
\providecommand\bcdotm{\mbox{\boldmath $\cdot$}}
\providecommand\rd{\rm d}
\providecommand\bdelt{\mbox{\boldmath $\delta$}}
\providecommand\bbet{\mbox{\boldmath $\beta$}}
\providecommand\bcdot{\boldsymbol{\cdot}}
\providecommand\bnab{\mbox{\boldmath $\nabla$}}
\providecommand\bsigma{\mbox{\boldmath $\sigma$}}
\providecommand\btimes{\mbox{\boldmath $\times$}}
\providecommand\bdelta{\mbox{\boldmath $\delta$}}
\providecommand\btau{\mbox{\boldmath $\tau$}}
\providecommand\bPhi{\mbox{\boldmath $\Phi$}}
\providecommand\bcdotm{\mbox{\boldmath $\cdot$}}
\providecommand\rd{\rm d} \providecommand\bdelt{\mbox{\boldmath
$\delta$}}
\newcommand{\revision}{{\color{blue}}}
\newcommand{\responseAB}{{\color{green}}}

\newcommand{\xtext}[1]{\mbox{#1}}

\title{Kinetic Density Functional Theory of Freezing}

\author{Arvind Baskaran}
 \email{baskaran@math.uci.edu}
\altaffiliation{Department of Mathematics, University of California Irvine, Irvine, CA 92697-3875.} 
 
\author{Aparna Baskaran}
 \email{aparna@brandeis.edu}
\altaffiliation{Martin Fisher School of Physics, Brandeis University, Waltham,
MA, USA.}

\author{John Lowengrub}
 \email{lowengrb@math.uci.edu}
 \altaffiliation{Department of Mathematics, University of California Irvine, Irvine, CA 92697-3875.} 


\begin{abstract}
A theory of freezing of a dense hard sphere gas is presented.
Starting from a revised Enskog theory, hydrodynamic equations that account for  non-local
variations in the density 
but local variations in the flow field are derived using a modified Chapman Enskog procedure. 
These hydrodynamic equations, which retain structural correlations, 
are shown to be effectively a time dependent density functional theory.
The ability of this theory to capture the solid liquid phase 
transition is established through analysis and numerical simulations.
\end{abstract}

\maketitle

Crystallization and melting are crucial in diverse contexts, ranging from crystal growth in manufacturing 
of semiconductor devices to the freezing process in ice cream making.
The equilibrium theory of  solid-liquid phase transitions is well developed \cite{Hansen2006,Lowen1994}.
However there are many non-equilibrium processes that are not yet well understood.
One such process is the melt flow interaction and its effect on the phase transition.
There exists a vast body of literature dedicated to understanding specific aspects of the effect of melt flow including
the work of Bradsley \cite{Bradsley1979}, Hurle \cite{Hurle1977}, Solan \& Ostrach \cite{Solan1979},
Pimputkar \& Ostrach \cite{Pimputkar1981} and Glicksman et al \cite{Glicksman1986}.
However a theory capable of predicting the consequences of the microscopic interactions in the system
and capturing the nanoscale details such as the lattice structure of the solid is still lacking.
It is our aim to develop such a theory.

Classical density functional theory (CDFT) introduced by Ramakrishnan and Youssouf  \cite{Ramakrishnan1979} (RY) 
and Haymet and Oxtoby \cite{Haymet1981} 
has been very successful in characterizing the equilibrium properties of the phase transition.
There has been a lot of progress in the development of reliable density functional theories to understand
solid liquid phase transitions (see \cite{Hansen2006}).
CDFT describes the freezing transition with respect to the one particle density field (the spatial probability distribution of particles)
at equilibrium.
The density at equilibrium is the minimizer of the free energy which in turn is a functional of the density.
This extremum principle describes the phase transition via the equilibrium particle density field which undergoes a transition
from a homogeneous (disordered phase or liquid) to an inhomogeneous (ordered phase of solid field).
The main challenge in developing a CDFT is the construction of a reliable free energy functional and
considerable progress has been made in this regard  \cite{Ramakrishnan1979,Haymet1981,Colot1985,Curtin1985,Curtin1986,Denton1989,Baus1990,Lutsko1990}.
However the CDFT approach is limited to describing the equilibrium states of the system determined by locating the local extrema and
saddle points of the free energy functional.
A description of approach to equilibrium is beyond the scope of this theory.

In recent years efforts have been focussed on the development of a time dependent Dynamic Density Functional Theory (DDFT)  \cite{Marconi1999, Yoshimori2005, Espanol2009, 2012Kaliadasis, Archer2009, Chavanis, Lutsko2012}.
These approaches aim to characterize the approach to equilibrium of a system of interacting particles close to equilibrium.
A direct consequence of the extremum principle is that the equilibrium density field is 
determined completely by the mean field interaction of the particles.
The mean field interaction at equilibrium is in turn known from the free energy functional if a reliable CDFT is available.
Taking advantage of this, Marconi and Tarazona\cite{Marconi1999} proposed 
that the density field can be time evolved with a mass flux driven by the mean field interaction.
This can be justified by means of a local equilibrium approximation.
The driving force in DDFT is in general the mean field force.
Of particular interest are the works Archer\cite{Archer2009}, Chavanis \cite{Chavanis} and Lutsko \cite{Lutsko2012}.
These authors  adopt a strategy of deriving a time dependent DDFT in the form of
hydrodynamics in which the free the energy enters the theory through a local equilibrium description 
for a non-local pressure.
However in these works \cite{Marconi1999, Yoshimori2005, Espanol2009, 2012Kaliadasis, Archer2009, Lutsko2012} the system of
interest was comprised of colloidal particles suspended in a solvent rather than a dense gas.
The drag force from the solvent makes the micro scale dynamics dissipative and drives the system rapidly to equilibrium.
This renders the system over-damped and justifies local equilibrium approximations at the level of the hydrodynamics.
However a dense gas of interacting particles considered here poses a challenge 
in that the dissipative processes that lead to equilibration must also be extracted from the inter-particle interactions.
This is done by means of a local equilibrium approximation at the level of  microscopic distribution functions rather than macroscopic fields (such as the density).
In particular this forms the basis of the Revised Enskog Theory\cite{VanBeijeren1973} (RET) where 
the local equilibrium approximation is used  to represent the two particle distribution as a
functional of one particle distributions and the local radial distribution function.
This will be the basis of our work.

Our aim is to develop a theory  to study the crystallization kinetics of a dense gas.
We use well established techniques in statistical mechanics to start from 
an appropriate kinetic theory 
and derive hydrodynamic equations for a dense gas close to the freezing transition.
This does not in anyway spare us from introducing a local equilibrium approximation.
The approximation is introduced at the level of the reduced distribution functions by using a closure 
relation to obtain a kinetic theory (the RET).
Then hydrodynamics can be derived self consistently through a generalized Chapman-Enskog procedure.
Local hydrodynamic equations for the RET have been previously derived\cite{VanBeijeren1973}.
In this approach\cite{VanBeijeren1973} the nonlocal collision operator is localized by gradient expansions of 
the non locality.
Kirkpatrick et al\cite{Kirkpatrick1990} noted the connection of the RET to DFT and derived non-local hydrodynamics
for the linearized collision operator.
To the best of our knowledge a non-local hydrodynamic description has not been previously derived from 
the RET.
The main contributions of this paper are the use of a generalized Chapman-Enskog procedure to 
derive such a  description, and show that it is effectively  a time dependent DFT and present numerical simulations that
confirm that the nonlocal hydrodynamic theory captures the solid-liquid phase transitions.

The rest of the paper is structured as follows.
In section  \ref{sec:kinetic_theory} we explain the procedure of  deriving macroscopic equations of motion from microscopic equations.
Here we outline the need for a kinetic theory, motivate and then present a simple derivation of the 
Revised Enskog Theory.
In section \ref{sec:hydrodynamics} we derive non-local hydrodynamics for the Revised Enskog theory.
The connections between the nonlocal hydrodynamic description and density functional theory are explored in section \ref{sec:KDFT}.
Finally some numerical results demonstrating the ability of the theory to predict the freezing transition
and capture the interactions between the crystal and the melt flow are presented in section \ref{sec:Numerics}.

\section{Microdynamics and Formal Non-equilibrium Statistical Mechanics}
\label{sec:kinetic_theory}

The goal of the theoretical program here is
to develop a framework that will allow us to understand the influence of
flow on freezing kinetics. Let us
 begin by considering the microdynamics of a system of $N$ identical
particles of mass $m$ with positions $\mathbf{r}^{N}=\{\mathbf{r}_{1}(t),%
\mathbf{r}_{2}(t)\ldots ,\mathbf{r}_{N}(t)\}$ and velocities $\mathbf{v}%
^{N}=\{\mathbf{v}_{1}(t),\mathbf{v}_{2}(t),\ldots ,\mathbf{v}_{N}(t)\}$ as a
function of time $t$. The dynamics of these particles is governed by the
Hamiltonian $H_{N}:=U+K+U^{ext}$
consists of the interparticle interaction energy $U:=\sum_{i,j}\mathcal{V}(%
\mathbf{r}_{i}(t),\mathbf{r}_{j}(t))$, the kinetic energy $K:=\sum_{i=1}^{N}%
\frac{1}{2}m\mathbf{v}_{i}^{2}$ and an external potential $%
U^{ext}:=\sum_{i=1}^{N}V^{ext}(\mathbf{r}_{i}).$ 
The equations of motion are :
\begin{eqnarray}
\partial _{t}\mathbf{r}_{i} &=&\mathbf{v}_{i},  \label{Eqn:Newton1} \\
\partial _{t}\mathbf{v}_{i} &=&-\frac{1}{m}\nabla _{\mathbf{r_{i}}}V(\br_i,t)-\frac{1%
}{m}\nabla _{\mathbf{r_{i}}}V^{ext}(\br_i),\label{Eqn:Newton2} 
\end{eqnarray}%
where $V$ is the potential energy associated with pairwise interaction of
the $i^{th}$ particle with the rest of the system. In the following, the
interaction potential is assumed to be pairwise additive, i.e., $V(\mathbf{r}%
_{i},t)=\sum_{j}\mathcal{V}(\mathbf{r}_{i}(t),\mathbf{r}_{j}(t)),$ and
radially symmetric i.e, $\mathcal{V}(\mathbf{r}_{i}(t),\mathbf{r}_{j}(t))=%
\mathcal{V}(\mid \mathbf{r}_{i}(t)-\mathbf{r}_{j}(t)\mid )$ . The function $V^{ext}$ is
an external potential such as gravity which will be taken as $V^{ext}=0$ for
simplicity. 
When the known initial condition is a macrostate rather than a microstate,
the equations of motion are most useful when expressed in terms of the phase
space probability distribution function $f^{(N)}(\mathbf{r}^{N},\mathbf{v}%
^{N},t)$. This function measures the probability of finding the system in
state $(\mathbf{r}^{N},\mathbf{v}^{N})$ at time t. The equation of motion in
terms of $f^{(N)}$ is given by the Liouville equation \cite{Hansen2006} :
\begin{equation}
\partial _{t}f^{(N)}+\sum_{i}\mathbf{v}_{i}\cdot \nabla _{\mathbf{r_{i}}%
}f^{(N)}+\frac{1}{m}\sum_{i}\nabla _{\mathbf{r_{i}}}V\cdot \nabla _{\mathbf{v%
}_{i}}f^{(N)}=0.  \label{Eqn:Liouville}
\end{equation}%
Since most observables of interest are sums of one particle and two particle
functions, it is useful to introduce reduced distribution functions $f^{(n)}$
defined as
\begin{equation}
f^{(n)}(\mathbf{r}^{n},\mathbf{v}^{n},t):=\frac{N!}{(N-n)!}\int d\mathbf{r}%
^{(N-n)}\int d\mathbf{v}^{(N-n)}f^{(N)}(\mathbf{r}^{N},\mathbf{v}^{N},t).
\end{equation}%
The time evolution of each reduced distribution function is obtained by
integrating the Liouville equation with respect to $\mathbf{r}_{n+1},\dots ,%
\mathbf{r}_{N}$ and $\mathbf{v}_{n+1},\ldots ,\mathbf{v}_{N}$. This gives us
a hierarchy of equations for the reduced distribution functions, known as the
Bogolyubov-Born-Green-Kirkwood-Yvon (BBGKY) hierarchy of equations \cite%
{Hansen2006}. The one particle distribution $f^{(1)}(\mathbf{r}_{1},\mathbf{v%
}_{1},t)$ gives the probability of finding a particle at $\mathbf{r}_{1}$
with velocity $\mathbf{v}_{1}$ at time $t$. The time evolution of this
function is given by the first equation of the BBGKY hierarchy :
\begin{equation}
\left( \partial _{t}+\mathbf{v}_{1}\cdot \nabla _{\mathbf{r_{1}}}\right)
f^{(1)}(\mathbf{r}_{1},\mathbf{v}_{1},t)=-\Omega ^{(12)}[f^{(2)}](\mathbf{r}%
_{1},\mathbf{v}_{1},\mathbf{r}_{2},\mathbf{v}_{2},t),  \label{Eqn:BBGKY1}
\end{equation}%
where 
\begin{equation}
\Omega ^{(12)}[f^{(2)}](\mathbf{r}_{1},\mathbf{v}_{1},\mathbf{r}_{2},%
\mathbf{v}_{2},t):=\int d\mathbf{r}_{2}\int d\mathbf{v}_{2}\frac{F_{12}}{m}%
\cdot \nabla _{\mathbf{v}_{1}}f^{(2)}(\mathbf{r}_{1},\mathbf{v}_{1},\mathbf{r%
}_{2},\mathbf{v}_{2},t),
\label{Eqn:Collision}
\end{equation}
\begin{equation}
F_{12}=-\nabla _{\mathbf{r_{1}}}\mathcal{V}(\mid
\mathbf{r}_{1}-\mathbf{r}_{2}\mid )
\end{equation}
and $f^{(2)}$ is the two particle
distribution function.

The
density, macroscopic velocity and temperature are naturally defined as :
\begin{equation}
\rho (\mathbf{r}_{1},t):=\int d\mathbf{v}_{1}f^{(1)}(\mathbf{r}_{1},\mathbf{v%
}_{1},t),
\end{equation}%
\begin{equation}
\rho(\mathbf{r}_{1},t) \mathbf{u}(\mathbf{r}_{1},t):=\int d\mathbf{v}_{1}\mathbf{v}%
_{1}f^{(1)}(\mathbf{r}_{1},\mathbf{v}_{1},t),
\end{equation}%
and
\begin{equation}
\frac{3}{2}\rho (\mathbf{r}_{1},t)k_B T (\br_1,t):=\int d\mathbf{v}_{1} \frac{1}{2} m \mid \bv_1 - \bu \mid^2f^{(1)}(\mathbf{r}_{1},\mathbf{v%
}_{1},t).
\end{equation}%

Suppose for simplicity the system is immersed in a heat bath, i.e., is isothermal at a temperature T, 
 the time evolution equations for these macroscopic variables are
simply the moments of   Eq. (\ref{Eqn:BBGKY1})
\begin{equation}
\begin{array}{c}
\partial _{t}\rho +\nabla _{\mathbf{r_{1}}}\cdot \rho \mathbf{u}=0, \\ \\
\partial _{t}(\rho \mathbf{u})+\nabla _{\mathbf{r_{1}}}\cdot (\rho \mathbf{u}%
 \mathbf{u}+\mathcal{P^K})=\tilde{\mathbf{J}}, \\ \\
\partial_{t} T = 0,
\end{array}
\label{Eq:Conserv_law}
\end{equation}%
where we have introduced the notation,
\begin{equation}
\mathcal{P^K}:=\int d\mathbf{v}_{1} (\mathbf{v}_{1}-\mathbf{u}) (\mathbf{v}_{1}-\mathbf{%
u})f^{(1)}(\mathbf{r}_{1},\mathbf{v}_{1},t)
\label{P_kinetic}
\end{equation}%
for the kinetic contribution to the static pressure and
\begin{equation}
\tilde{\mathbf{J}}:=\int d\mathbf{v}_{1}\mathbf{v}_{1}\Omega ^{(12)}[f^{(2)}](%
\mathbf{r}_{1},\mathbf{v}_{1},\mathbf{r}_{2},\mathbf{v}_{2},t)  \label{J}
\end{equation}%
is the contribution from the inter-particle interactions. Also, $\mathbf{u}%
\mathbf{u}$ is a rank 3 tensor such that $\left( {\mathbf{u}
\mathbf{u}}\right) _{ij}=u_{i}u_{j}$. If the above equations can be closed,
i.e., $\mathcal{P}^K$ and $\tilde{\mathbf{J}}$ can be expressed as functionals of the
fields $\rho $ and $\mathbf{u}$, we obtain a
macroscopic description of the hydrodynamics of the system.
The approach developed in this paper can be naturally extended to account for temperature variations,
the release of latent heat and convective instabilities that arise.
This will be considered in a future work.

\subsection{Hydrodynamics for Over-damped Systems} \label{subsection_overdamped}

Before we outline our theoretical framework for deriving the hydrodynamic
description, it is important to note that there are several other routes that one
may take to estimate the momentum fluxes if one were considering an
over-damped system like a system of colloidal particles. The particles
experience a drag force as they move through the solvent that damps out the
thermal fluctuations and drives the system to equilibrium. In this case Eq. (\ref{Eqn:Newton1}) and
Eq. (\ref{Eqn:Newton2}) are replaced by :
\begin{eqnarray}
\partial _{t}\mathbf{r}_{i} &=&\mathbf{v}_{i}  \label{Eqn:Newton_colloid} \\
\partial _{t}\mathbf{v}_{i} &=&-\frac{1}{m}\nabla _{\mathbf{r_{i}}}V-\frac{1%
}{m}\nabla _{\mathbf{r_{i}}}V^{ext}-\nu \mathbf{v}_{i}
\end{eqnarray}%
where $\nu $ is a positive constant. There are at least two
routes to obtaining a closed form for the tensor $\mathcal{P}^K$ and the
vector $\tilde{\mathbf{J}}$ in Eq. (\ref{Eq:Conserv_law}) for the over-damped particle system :

\begin{enumerate}
\item One approach is to try to characterize the unknown momentum flux as
the gradient of a scalar pressure. This would correspond to neglecting all
dissipative processes coming from inter-particle interactions. The
hydrodynamics in this case takes the form
\begin{equation}
\begin{array}{c}
\partial _{t}\rho +\nabla _{\mathbf{r_{1}}}\cdot \rho \mathbf{u}=0 \\
\partial _{t}(\rho \mathbf{u})+\nabla _{\mathbf{r_{1}}}\cdot (\rho \mathbf{u} \mathbf{u})+\nu \rho \mathbf{u}+\nabla _{\mathbf{r_{1}}}p=0%
\end{array}%
\end{equation}%
where $p(\mathbf{r}_{1},t)$ is the pressure. This pressure for a
non-interacting ideal gas at equilibrium is simply $p=\rho k_{B}T$. For a
system with pair potential interactions we have the equilibrium relation $Nd\mu
=-SdT+Vdp $, where $\mu $ is the chemical potential. For the isothermal case
this is simply $\rho d\mu =dp$. Assuming local equilibrium this chemical
potential can be defined by appealing to density functional theory for a
dense gas as $\mu :=\frac{\delta \mathcal{F}[\rho ]}{\delta \rho }$, where $%
\mathcal{F}[\rho ]$ is the intrinsic Helmholtz free energy as a functional
of the density field. This gives the hydrodynamic equations:
\begin{equation}
\begin{array}{c}
\partial _{t}\rho +\nabla _{\mathbf{r_{1}}}\cdot \rho \mathbf{u}=0, \\
\partial _{t}(\rho \mathbf{u})+\nabla _{\mathbf{r_{1}}}\cdot (\rho \mathbf{u} \mathbf{u})
+\nu \rho \mathbf{u}=-\rho \nabla _{\mathbf{r_{1}}}\frac{%
\delta \mathcal{F}[\rho ]}{\delta \rho },%
\end{array}%
\label{Eqn:DDFT_1}
\end{equation}%
which is the approach used by Lutsko \cite{Lutsko2012}.
Appealing to CDFT (see \cite{Hansen2006})  we have $\rho \Grad \frac{\mathcal{F}[\rho ]}{\delta \rho } = \frac{k_B T}{m} \left(\Grad  (\rho) +\rho \Grad C^{(1)} [ \br_1 \mid \rho] + \rho \Grad V^{ext} ( \br_1) \right)$,
where $C^{(1)}$ is the direct correlation function.
The direct correlation function is simply the mean field external potential
that produces the same equilibrium structure in a non-interacting fluid as
that of the interactions.
Thus the hydrodynamic model in Eq. (\ref{Eqn:DDFT_1}) is simply a non-interacting fluid 
driven by the mean field force of the interactions. 
\item The second approach is to evaluate all fluxes assuming the local
equilibrium distribution is $f^{(1)} ( \mathbf{r}_{1}, \mathbf{v}_{1},t ) =
\rho(\mathbf{r}_1) \phi^M (\mathbf{r}_1,\mathbf{v}_1,t)$, the local
Maxwellian, where
\begin{equation}  \label{Eqn:Maxwellian}
\phi^M( \mathbf{r}_1,\mathbf{v}_1,t ) = \left( \frac{m}{2\pi k_{B}T}\right)
^{3/2} e^{-m\left( \mathbf{v}_{1}-u\left( \mathbf{r}_1,t\right) \right)
^{2}/2k_{B}T}
\end{equation}
Now the term $\nabla_{\mathbf{r_1}} \cdot \mathcal{P}$ reduces to $\nabla_{%
\mathbf{r_1}} \rho k_B T$. Then to evaluate the inter-particle interactions
Archer \cite{Archer2009} proposed that the forces can be replaced by the
mean field force to obtain the same model in the previous case (also see
Marconi et al \cite{Marconi1999}).
In a similar manner Chavanis \cite{Chavanis} considered thermostated Brownian
particles within the framework of Smoluchowski equations and 
approximated the interactions by the mean field force (as done by Archer \cite{Archer2009})
to obtain a model similar to the one in Eq. (\ref{Eqn:DDFT_1}).
\end{enumerate}

The common ingredient in both approaches is to assume that the system tends to equilibrate and to
make a local equilibrium approximation. In the context of colloidal
particles in a suspension, a local equilibrium assumption is justified, as
the motion of the particles is damped by the friction force they experience
from the solvent. This drives the system to equilibrium even when one
neglects dissipative processes that arise from the inter-particle
interactions of the colloids. However our interest is in characterizing the
effect of melt flow on the crystallization kinetics of a dense gas and not a colloidal suspension. 
Therefore, the macroscopic
description should capture structural information as in the closures
mentioned above, but at the same time capture dissipation and equilibration
as well. In the non-equilibrium statistical mechanics framework above, the
momentum fluxes are given in terms of moments of the solution to the
Liouville equation. However, the Liouville equation has time reversal symmetry. In
order to be able to derive dissipative hydrodynamics for this system one
must break the time reversal symmetry by appealing to the techniques of
non-equilibrium statistical mechanics and obtain a kinetic theory that
satisfies an H-Theorem and thereby guarantees equilibration \cite{Resibois1965}. 
In the case of an isothermal system the system will dissipate momentum and equilibrate
to a steady state with a constant velocity field.
Once an H-Theorem has been established, one can expand the system about an
equilibrium or a local equilibrium solution to derive hydrodynamics in a
systematic manner. This is done formally through the
Chapman-Enskog method. Starting from an appropriate kinetic theory and
deriving a hydrodynamic model that captures structural correlations that can
be used to study the phase transition and the effect of melt flow
on crystallization kinetics is one of the goals in this paper.



\section{Kinetic Theory and Non-local Hydrodynamics}
\label{sec:hydrodynamics}

Let us begin by considering the first equation in the BBGKY hierarchy Eq. (\ref{Eqn:BBGKY1}).
The first step in developing a kinetic theory is
formulating a closure ansatz by representing the interaction term $\Omega^{12}$ as a
functional of $f^{(1)}$ :
\begin{equation}
\Omega ^{(12)}[f^{(2)}](\mathbf{r}_{1},\mathbf{v}_{1},\mathbf{r}_{2},\mathbf{%
v}_{2},t)=\Omega _{kinetic}^{(12)}[f^{(1)},f^{(1)}]((\mathbf{r}_{1},\mathbf{v%
}_{1},\mathbf{r}_{2},\mathbf{v}_{2},t)
\end{equation}
If we choose the inter-particle potential to be of the simplest form that
undergoes a freezing transition, i.e a hard sphere gas, Eq. (\ref{Eqn:Collision})
can be rewritten as\cite{Resibois1965} :
\begin{equation}
\begin{array}{rl}
\Omega _{HS}^{(12)}[f^{(2)}(\mathbf{r}_{1},\mathbf{v}_{1},\mathbf{r}_{2},%
\mathbf{v}_{2},t)] & =\displaystyle\int d\mathbf{v}_{2}d\mathbf{r}_{12}\Theta (\hat{\mathbf{g%
}}\bcdot\mathbf{\hat{r}}_{12})\mid \mathbf{g}\bcdot\mathbf{\hat{r}}_{12}\mid \\%
& \qquad \quad \left[ \delta (\mathbf{r}_{12}-\bsigma)\hat{b}^{-1} 
- \delta (\mathbf{r}_{12}+%
\bsigma)\right] f^{(2)}(\mathbf{r}_{1},\mathbf{v}_{1},\mathbf{r}_{2},\mathbf{%
v}_{2},t).  
\end{array}
\label{Eqn: BBGKY_HS}
\end{equation}%
where $\mathbf{r}_{12}=\mathbf{r}_{1}-\mathbf{r}_{2}$, $\mathbf{g}=(\mathbf{v%
}_{1}-\mathbf{v}_{2})$, $\Theta(x)$ is the Heaviside step function  and $\hat{b}^{-1}$ is the operator that maps $(%
\mathbf{v}_{1},\mathbf{v}_{2})$ to the restituting velocities,

\begin{equation}
\begin{array}{c}
\mathbf{v}_{1}^{\prime }=\hat{b}\mathbf{v}_{1}=\mathbf{v}_{1}-\hat{\bsigma}(%
\mathbf{g}\cdot \hat{\bsigma}) \\
\mathbf{v}_{2}^{\prime }=\hat{b}\mathbf{v}_{2}=\mathbf{v}_{2}+\hat{\bsigma}(%
\mathbf{g}\cdot \hat{\bsigma})%
\end{array}%
\end{equation}%
where $\bsigma=\sigma \hat{\bsigma}$ ( $=\mathbf{r}_{1}-\mathbf{r}_{2}$ at
contact of two hard spheres ) with $\sigma $ being the hard sphere radius
 and $\hat{\bsigma}$  a unit vector normal to the point of contact of the two spheres.
In order to close the hierarchy we appeal to a local equilibrium
approximation. At equilibrium we know that
\begin{equation}
f_{eq}^{(2)}(\mathbf{r}_{1},\mathbf{v}_{1},\mathbf{r}_{2},\mathbf{v}%
_{2})=G_{2}[\mathbf{r}_{1},\mathbf{r}_{2}\mid \rho _{eq}]f^{(1)}(1)f^{(1)}(2),
\label{Eq:local_equi}
\end{equation}%
where $f^{(1)}(1)=f_{eq}^{(1)}(\mathbf{r}_{1},\mathbf{v}%
_{1}),f^{(1)}(2)=f_{eq}^{(1)}(\mathbf{r}_{2},\mathbf{v}_{2})$ and $G_{2}[%
\mathbf{r}_{1},\mathbf{r}_{2}\mid \rho ]$ is the pair distribution function
as a functional of the local density field $\rho $. 
Assuming that the pre-collision distribution of particles satisfies
Eq. (\ref{Eq:local_equi}) one can derive the Revised Enskog Theory (RET)
 introduced
by Van Beijeren and Ernst \cite{VanBeijeren1973} (also see Lutsko \cite{Lutsko1996,Lutsko2001}). 
The resulting kinetic theory takes the form
\begin{equation}
\partial _{t}f(\mathbf{r}_{1},\mathbf{v}_{1},t)+\mathbf{v}_{1}\cdot \nabla _{%
\mathbf{r_{1}}}f(\mathbf{r}_{1},\mathbf{v}_{1},t)=J_{E}(G_{2}[\mathbf{r}_{1},%
\mathbf{r}_{2}\mid \rho ],f(1),f(2)),  \label{RET}
\end{equation}%
where
\begin{equation}
\begin{array}{rl}
J_{E}(G_{2}[\mathbf{r}_{1},\mathbf{r}_{2} \mid \rho ],f(1),f(2)) &= \displaystyle \int d%
\mathbf{v}_{2} d\hat{\bsigma}\sigma ^{2}\Theta (\hat{\mathbf{g}}\bcdot\hat{%
\sigma})\mid \mathbf{g}\bcdot\hat{\bsigma}\mid 
\\ & \qquad \left(G_{2}[\mathbf{r}_1,\mathbf{r}_1-\bsigma \mid \rho (t)] f(\mathbf{r}_{1},\mathbf{v}_{1}^{\prime },t)f(\mathbf{r}_{1}-
\bsigma,\mathbf{v}_{2}^{\prime },t) \right.
\notag \\
& \qquad \left. -G_{2}[\mathbf{r}_1,\mathbf{r}_1+\bsigma \mid \rho (t)]f(\mathbf{r}_{1},\mathbf{v}_{1},t)f(%
\mathbf{r}_{1}+\bsigma,\mathbf{v}_{2},t)\right).  
\end{array}
\label{RETColl}
\end{equation}
The RET  breaks time reversal invariance and
captures dissipation. An H-theorem was proved for this system by Resibois
\cite{Resibois1978} (also see Piasecki \cite{Piasecki1987}). This allows one
to expect equilibration in the long time limit to the Maxwell Boltzmann
Distribution and to perform a Chapman-Enskog-like expansion to derive a
macroscopic hydrodynamic description. This has been done extensively in the
context of local hydrodynamic theories for the fluid phase (see Resibois and DeLeener\cite{Resibois1965}).
Here, we
generalize this framework to derive non-local hydrodynamics that captures
 structural information in the fluid (and solid) as well. Also, it is worth comparing the local
equilibrium approximation suggested for the over-damped system outlined in section (\ref{subsection_overdamped}) 
with the one
used in deriving the RET. The difference is that the derivation of the RET
merely imposes the long range structure of the liquid $G_{2}$ (see Eq. (\ref%
{Eq:local_equi})) that is needed for the freezing transition at equilibrium
allowing the non-equilibrium distribution to be determined as a consequence.
The over damped approach determines the local equilibrium mean field force ($\Grad C^{(1)}[\br_1|\rho]$) to
derive macroscopic equations.



Now we are ready to derive the hydrodynamic equations using the RET in Eq. (\ref%
{RET}) as our starting point. The first step is to assume that on the length
and time scales of interest, the RET admits a normal solution of the form
\begin{equation}
f(\mathbf{r}_{1},\mathbf{v}_{1},t)=f_{norm}(\mathbf{v}_{1}\mid \rho (\mathbf{%
r}_{1},t),\mathbf{u}(\mathbf{r}_{1},t)).  \label{Eqn:functional}
\end{equation}%
Here the space and time dependence of the distribution are implicit
through the functional dependence on the macroscopic variables. 
We seek to derive a hydrodynamic description that retains nonlocal information in
the density field while being local in the velocity field alone.
This is accomplished through a gradient expansion of the velocity field.
In anticipation of such an expansion we propose the following ansatz
\begin{equation}
f_{norm}( \bv_1 \mid \rho ( \br_1,t), \bu( \br_1,t ) ) = \rho (\br_1, t) \phi ( \bv_1 | \bu ( \br_1,t) ).
\label{F_ansatz}
\end{equation}
When this form is substituted back into the RET, the space and time derivatives occur
only through the functional dependence on hydrodynamic fields. 
The macroscopic balance equations for the density $\rho$ and the momentum $\rho \bu$ 
 with the ansatz in Eq. (\ref{F_ansatz}) take the form
\begin{equation}
\begin{array}{c}
\partial_t \rho + \Div (\rho \bu) = 0,\\
\partial_t \rho \bu + \Div ( \rho \bu \bu ) + \Div \mathcal{P} ={\bf  J},
\end{array}
\label{RET_macro_balance}
\end{equation}
where the pressure tensor $\mathcal{P}(\br_1,t)$  has both a kinetic and a collisional transfer part, i.e $\mathcal{P} = \mathcal{P}^K + \mathcal{P}^C$.
The kinetic contribution to pressure $\mathcal{P}^K$ is defined in Eq. (\ref{P_kinetic}) and the collisional transfer contribution  is given by (see Appendix \ref{BALANCE_EQ_APPENDIX} for details)
\begin{equation}
\mathcal{P}^C = \frac{1}{2} \int d \bv_1 d \bv_2 d \hat{\bsigma} \sigma^3 \hat{\bsigma} \hat{\bsigma}( \bg \cdot \hat{\bsigma})^2 \Theta(\bg \cdot \hat{\bsigma})  \int d \lambda F (\br_1 - (1-\lambda) \bsigma, \bv_1, \br_1+ \lambda \bsigma, \bv_2 ). 
\label{P_collision}
\end{equation}
In the above equation 
\begin{equation}
F(\br_1, \bv_1, \br_2, \bv_2) =G_2[\br_1, \br_2| \rho] \rho (\br_1) \rho (\br_2) \phi(\br_1,\bv_1) \left( \phi( \br_2, \bv_2) - \phi(\br_1, \bv_2) \right).
\label{eq_def_F}
\end{equation}
Finally the remaining collisional contribution is given by 
\begin{equation}
\label{Eqn:J}
{\bf J} = \displaystyle \frac{1}{2}\int d%
\bv_1d \mathbf{v}_{2}d\hat{\bsigma}\sigma ^{2}  \hat{\bsigma}  ( \hat{\bsigma} \cdot \bg)^2     \rho(\br_1)  G_{2}[\mathbf{r}_1,\mathbf{r}_1-\bsigma \mid \rho (t)] \rho( \br_1 -\bsigma) \phi(\br_1, \bv_1) \phi(\br_1, \bv_2) . 
\end{equation}
The standard Chapman-Enskog procedure (see Resibois and DeLeener\cite{Resibois1965})
 aims to construct normal solutions of different orders of 
gradients in all the macroscopic variables ($\rho$ and $\bu$).
However in order to derive hydrodynamics that are non-local in the density field,
we construct normal solutions at different orders in gradients of the velocity field alone. 
To this end, we
introduce a uniformity parameter $\varepsilon $ that measures the order in
gradients in the velocity field $\mathbf{u}$ and
we seek to construct a
normal solution of the form%
\begin{equation}
f_{norm}=f_{0}+\varepsilon f_{1}+\ldots = \rho ( \phi_0 + \varepsilon \phi_1 + \ldots ).
\label{eq_expansion}
\end{equation}%
This in turn induces an expansion in the collision operator (see Appendix \ref{CE_APPENDIX} for details)  and the time
derivative 
\begin{equation*}
J_{E}=J_{E}^{\left( 0\right) }+\varepsilon J_{E}^{\left( 1\right)
}+\ldots \quad \text{and} \quad \partial _{t}=\partial _{t}^{\left( 0\right) }+\varepsilon \partial
_{t}^{\left( 1\right) }+ \ldots .
\end{equation*}%
We use the macroscopic balance equations to eliminate the
time derivatives in favor of mass and momentum fluxes, which allows us to
construct a self-consistent normal solution perturbatively.

\subsection{Euler Order Hydrodynamics}

To lowest order in the uniformity parameter the kinetic equation Eq. (\ref{RET}) reduces to
\begin{equation}
\partial _{t}^{\left( 0\right) } (\rho \phi_{0} ) + (\bv_1 \cdot \Grad \rho) \phi_0  =J_{E}^{\left( 0\right)
}(G_{2},f_{0}\left( 1\right) ,f_{0}(2))
\end{equation}%
and the macroscopic conservation laws  Eq. (\ref{RET_macro_balance}) reduce to
\begin{equation*}
\partial_t^{(0)} \rho = - \bu \cdot \Grad \rho, \qquad \partial _{t}^{\left( 0\right) }\rho \mathbf{u} = - (\bu  \bu) \cdot \Grad \rho + {\bf J}_{0} - \Div {\mathcal{P}^{K}}^{(0)} , 
\end{equation*}%
where 
\begin{equation}
\label{Eqn:J0}
{\bf J}_0 = \displaystyle \frac{1}{2}\int d%
\bv_1d \mathbf{v}_{2}d\hat{\bsigma}\sigma ^{2}  \hat{\bsigma}  ( \hat{\bsigma} \cdot \bg)^2     \rho(\br_1)  G_{2}[\mathbf{r}_1,\mathbf{r}_1-\bsigma \mid \rho (t)] \rho( \br_1 -\bsigma) \phi_0(\br_1, \bv_1) \phi_0(\br_1, \bv_2) . 
\end{equation}
and ${\mathcal{P}^{K}}^{(0)} = \rho \int d \bv_1 (\bv_1 - \bu ) (\bv_1 -\bu) \phi_0 $. 
Using the conservation law to eliminate the time derivatives we have
\begin{equation}
\left(  ( \bv_1 - \bu ) \cdot \Grad \rho \right) \phi_0 + \left( \bar{\mathcal{P}^{K}}^{(0)}  \cdot \Grad \rho - \rho {\bf J}_0 \right) \cdot \nabla_{\bv_1} \phi_0 = J_{E}^{\left(
0\right) }(G_{2},f_{0}\left( 1\right) ,f_{0}(2)),
\label{Euler_integral_equation}
\end{equation}%
where $ \bar{\mathcal{P}^{K}}^{(0)} = \int d \bv_1 (\bv_1 - \bu )  ( \bv_1 - \bu ) \phi_0 $.
It is can be verified (see Appendix \ref{EULER_APPENDIX}) that a local Maxwellian velocity distribution of the form
\begin{equation}
\phi^{M}(\mathbf{r}_{1},\mathbf{v}_{1},t) =\left( \frac{m}{2\pi
k_{B}T}\right) ^{3/2} e^{-m\left( \mathbf{%
v}_{1}-\bu\left( \mathbf{r}_{1},t\right) \right) ^{2}/2k_{B}T},
\label{eqn:maxwellian}
\end{equation}%
indeed solves the above equation.
Using the Maxwellian we can easily evaluate
\begin{equation}
\mathbf{J}_{0}=\sigma ^{2}\left( \frac{k_{B}T}{m}\right) \rho \left( \mathbf{%
r}_{1}\right) \int d\hat{\bsigma}\hat{\bsigma}\rho \left( \mathbf{r}_{1}-%
\bsigma\right) G_{2}\left[ \mathbf{r}_{1},\mathbf{r}_{1}-\bsigma|\rho \right]
.  \label{J0}
\end{equation}%
The pressure flux to Euler order is also readily
evaluated to give $\Div \mathcal{P} = \Div {\mathcal{P}^{K}}^{(0)}=\nabla _{%
\mathbf{r_{1}}}(\frac{\rho k_{B}T}{m})$ . So, to this order in the
perturbation theory, the hydrodynamic equations take the form%
\begin{equation}
\begin{array}{rll}
\partial _{t}\rho +\nabla _{\mathbf{r_{1}}}\cdot \rho \mathbf{u} &=&0, \\
\partial _{t}(\rho \mathbf{u})+\nabla _{\mathbf{r_{1}}}\cdot (\rho \mathbf{u} \mathbf{u})+\nabla _{\mathbf{r_{1}}}\left( \frac{\rho k_{B}T}{m}%
\right) &=&\sigma ^{2}\left( \frac{k_{B}T}{m}\right) \rho \left( \mathbf{r}%
_{1}\right) \int d\hat{\bsigma}\hat{\bsigma}\rho \left( \mathbf{r}_{1}-\bsigma%
\right) G_{2}\left[ \mathbf{r}_{1},\mathbf{r}_{1}-\bsigma|\rho \right].
\end{array}
\label{eqn:Euler_eqn}
\end{equation}

\subsection{Navier-Stokes Order Hydrodynamics}

To obtain the viscous contributions, we consider the normal solution to order $\varepsilon $ in the
perturbation theory. From Eq. (\ref{RET}) to order $\varepsilon $, we have
\begin{equation}
\partial_t^{(0)} ( \rho \phi_1 ) + \partial_t^{(1)} ( \rho \phi_0 ) + \left( \bv_1 \cdot \Grad \rho \right) \phi_1 +\left( \bv_1 \cdot \Grad \phi_0 \right) \rho
= J^{(1)}_E[\rho \phi_0, \rho \phi_0]  + \mathcal{L}[\phi_1],
  \label{OrderEpsilon}
\end{equation}%
where $\mathcal{L}[\phi_1]:=J_{E}^{\left( 0\right) }\left[ \rho \phi_{0},\rho \phi_{1}\right]
+J_{E}^{\left( 0\right) }\left[ \rho \phi_{1},\rho \phi_{0}\right]$. 
Since the first three moments of the local Maxwellian are captured exactly we note that
\begin{equation*}
\int d \bv_1\left(
\begin{array}{c}
1 \\
\mathbf{v}_1 \\
\mid \mathbf{v}_1\mid^{2}%
\end{array}%
\right) \phi_{1}=0.
\end{equation*}
Using this, the macroscopic balance equations to Navier Stokes order can be simplified to
\begin{equation}
\partial_t^{(1)} \rho = -\rho ( \Div \bu) ,  \quad \partial_t^{(1)} (\rho \bu) = -\rho \Div (\bu \bu) + {\bf J}_1,
\label{Eq_NS_HD}
\end{equation}
where the components of ${\bf J}_1$ are given by (see Appendix \ref{APPENDIX_J} for details)
\begin{equation}
\label{Eqn:J1}
{\bf J}_{1k} = \mathcal{J}_{1ijk}  \left( \int d
\bv_1  (v_{1i}v_{1j} - \frac{1}{3} \delta_{ij} \mid \bv_1 \mid^2 )\phi_1(\br_1, \bv_1) \right) 
\end{equation}
and
\begin{equation}
\mathcal{J}_{1ijk} = \int d\hat{\bsigma}\sigma ^{2} \sigma_k   \sigma_i \sigma_j  \rho(\br_1)\rho( \br_1 -\bsigma)   G_{2}[\mathbf{r}_1,\mathbf{r}_1-\bsigma \mid \rho (t)]. 
\end{equation}

The time derivatives in Eq. (\ref{OrderEpsilon}) can now be eliminated in favor of spatial derivatives of the hydrodynamic fields
using Eq. (\ref{Eq_NS_HD}) to obtain  the integro-differential equation for $\phi_1$ :
\begin{equation}
\begin{array}{l}
 \displaystyle \mathcal{L}[ \phi_1 ] - ( (\bv_1- \bu) \cdot \Grad \rho) \phi_1 
 \displaystyle + \rho \nabla_{\bv_1} \phi_1 \cdot \left( {\bf J}_0 - \frac{k_B T}{m} \Grad \rho \right)   \\ \\
 \displaystyle \qquad + \left( (\bv_1 - \bu) \cdot  {\bf J}_1 \right)\frac{k_B T}{m} \phi_0\\ \\
 \qquad \qquad = 
 \displaystyle  - (\rho \Div \bu ) \phi_0   
  \displaystyle    - (\bv_1- \bu) \cdot ( (\bv_1 - \bu) \cdot \Grad \bu) \frac{k_B T}{m} \phi_0 \rho  -  \mathcal{K}[ \bv_1 \mid \rho] : \Grad \bu (\br_1,t)   
  \end{array}
\end{equation}
where the tensor $\mathcal{K}$ is given by (see Appendix \ref{APPENDIX_JE})
\begin{equation}
\begin{array}{rl}
\mathcal{K}_{ij}[ \bv_1 \mid \rho ] =  &  \displaystyle \int  d\bv_2 d\sigh \Theta( \gh \bcdot \sigh ) (\bg \bcdot \sigh)  \\
& \displaystyle \left( G[\br_1,\br_1+\bsigma | \rho] \rho (\br_1,t) \phi_0(\br_1,\bv_1' )  \phi_0(\br_1,\bv_2' ) \rho(\br_1+ \bsigma,t)  \frac{(\bv_2' - \bu)_i}{k_BT}  \sigma_j  \right. \\
 & \displaystyle + \left. G[\br_1,\br_1-\bsigma | \rho] \rho (\br_1,t)  \phi_0(\br_1,\bv_1)  \phi_0(\br_1,\bv_2 )  \rho(\br_1- \bsigma,t)    \frac{(\bv_2 - \bu)_i}{k_BT} \sigma_j \right).
 \end{array}
\end{equation}
It is easy to see that this admits a solution $\phi_1$ of the form
\begin{equation}
\phi_1 ( \bv_1 ) = \mathcal{C} [ \bv_1 | \rho] : \mathcal{D} + \mathcal{Q} [ \bv_1 | \rho] ( \Div \bu ).
\label{CE_FORM}
\end{equation}
where the symmetric stress tensor $\mathcal{D}$ is defined as
\begin{equation}
\mathcal{D}_{ij} := \frac{1}{2} \left ( \partial_i u_j+ \partial_j u_i - \frac{2}{3} \delta_{ij} \Div \bu \right). 
\end{equation}
The tensor $\mathcal{C}$ and the scalar $ \mathcal{Q}$ are in turn to be determined by substituting this form back into the
integro-differential equation to obtain integral equations for $\mathcal{C}$ and $ \mathcal{Q}$ given by :
\begin{equation}
\begin{array}{l}
 \displaystyle \mathcal{L}[ \mathcal{C}_{ij} ] - ( (\bv_1- \bu) \cdot \Grad \rho) \mathcal{C}_{ij} 
 \displaystyle + \rho \left( {\bf J}_0 - \frac{k_B T}{m} \Grad \rho \right) \cdot \nabla_{\bv_1} \mathcal{C}_{ij} \\ \\
 \displaystyle \qquad + \left( (\bv_1 - \bu)_k \mathcal{J}_{1lmk}  \left( \int d
\bv_2 d\hat{\bsigma} (v_{2l}v_{2m} - \frac{1}{3} \delta_{lm} \mid \bv_2 \mid^2 )\mathcal{C}_{ij}(\bv_2) \right) \right)\frac{k_B T}{m} \phi_0 \\ \\
 \qquad \qquad = 
  \displaystyle    - (\bv_1- \bu)_i  (\bv_1 - \bu)_j  \frac{k_B T}{m} \phi_0 \rho  -  \mathcal{K}_{ij} 
  \end{array}
\end{equation}
and
\begin{equation}
\begin{array}{l}
 \displaystyle \mathcal{L}[ \mathcal{Q} ] - ( (\bv_1- \bu) \cdot \Grad \rho) \mathcal{Q} 
 \displaystyle + \rho \left( {\bf J}_0 - \frac{k_B T}{m} \Grad \rho \right) \cdot \nabla_{\bv_1} \mathcal{Q} \\ \\
 \displaystyle \qquad  \left( (\bv_1 - \bu)_k \mathcal{J}_{1ijk}  \left( \int d
\bv_2 d\hat{\bsigma} (v_{2i}v_{2j} - \frac{1}{3} \delta_{ij} \mid \bv_2 \mid^2 )\mathcal{Q}(\bv_2) \right) \right)\frac{k_B T}{m} \phi_0 \\ \\
 \qquad \qquad = 
  \displaystyle  - \rho \phi_0   - \frac{1}{3} \mid\bv_1- \bu\mid^2 \frac{k_B T}{m} \phi_0 \rho  -  \frac{1}{3} Tr \left[\mathcal{K} \right].
  \end{array}
\end{equation}
Once these equations are solved $\phi_1$ can be used to obtain the Navier-Stokes equations as shown in 
Appendices \ref{APPENDIX_KINETIC} ,\ref{APPENDIX_COLLISIONAL_TRANSFER} and \ref{APPENDIX_J} to get :
\begin{equation}
\begin{array}{ll}
\displaystyle\partial _{t}\rho +\nabla _{\mathbf{r_{1}}}\cdot \rho \mathbf{u}
& =0, \\
\displaystyle\partial _{t}(\rho \mathbf{u})+\nabla _{\mathbf{r_{1}}}\cdot
(\rho \mathbf{u} \mathbf{u}) + \Grad \left( \frac{\rho k_B T}{m} \right) & = \displaystyle  \mathbf{J}_{0} +  \mathbf{J}_{1} - \nabla_{\br_1} \cdot{ \mathcal{P}^{K}}^{(1)} 
- \nabla_{\br_1} \cdot {\mathcal{P}^C}^{(1)},%
\end{array}
\label{NS_FULL}
\end{equation}%
where $\mathbf{J}_{0}$ is given in Eq. (\ref{J0}), and the dissipative terms are 
given by
\begin{equation}
{\bf J}_{1k} =  \mathcal{J}_{1ijk} (\mu^{K}_{ijlm} \mathcal{D}_{lm} + \nu^K_{ij} \Div \bu), 
\end{equation}
\begin{equation}
{\mathcal{P}^{K}}^{(1)}_{ij} = \mu^{K}_{ijlm} \mathcal{D}_{lm} + \nu^K_{ij} \Div \bu,
\end{equation}
and 
\begin{equation}
{\mathcal{P}^C}^{(1)}_{ij} = \mu_{lmij} \partial_l u_m,
\end{equation}
and the transport coefficients are given by
\begin{equation}
\mu^{K}_{ijlm} =\rho   \int d \bv_1 (v_{1i} v_{1j} -\frac{1}{3} \delta_{ij} \mid \bv_1 \mid^2)   \mathcal{C}_{lm} [ \bv_1],
\end{equation} 
\begin{equation}
\nu^{K}_{ij} = \rho \int d \bv_1 (v_{1i} v_{1j} -\frac{1}{3} \delta_{ij} \mid \bv_1 \mid^2)   \mathcal{Q}[ \bv_1],
\end{equation}
and
\begin{equation}
\mu_{ijkl}  =  4 \left( \frac{m} { \pi k_B T}\right)^{1/2} \int  d \hat{\bsigma} \sigma^3 \sigma_i \sigma_j \sigma_k \sigma_l   \int_0^1 d \lambda G_2 [ \br_1 -(1-\lambda)\bsigma, \br_1 + \lambda \bsigma ] \rho (\br_1 +\lambda \bsigma) \rho(\br_1 -(1-\lambda)\bsigma). 
\end{equation}

The hydrodynamic equations derived in Eq. (\ref{NS_FULL}) account for the non-local variations in density and 
local variations in the flow field.
The unknown quantities so far are the pair distribution function $G[\br_1, \br_2 \mid \rho]$ and the solutions to the integro-differential equation 
$\mathcal{C}[\bv_1 \mid \rho]$ and $\mathcal{Q}[\bv_1 \mid \rho]$.
Given a functional form for $G$ one can solve for $\mathcal{C}$ and $\mathcal{Q}$ by using a polynomial basis representation such as a Sonine polynomial basis (see Resibois \cite{Resibois1965}).
Determining the transport coefficients ( $\mathcal{C}$ and $\mathcal{Q}$ ) is interesting and worthy of pursuit but we defer this for future work.

We note that the hydrodynamic equations derived above have the ideal gas static pressure ${\mathcal{P}^K}^{(0)}$ as in the case of the conventional Navier Stokes equations (see Resibois \cite{Resibois1965}). 
However the viscous or dissipative terms in the new hydrodynamic equations are different and non-local.
It is also striking that the dissipative processes appear in the Euler equations.
At Euler order the macroscopic balance equation for the momentum is no longer a local conservation law.

It is easy to see the critical points $\rho_{eq}$ of the Helmholtz free energy
functional for the hard sphere system are stationary solutions ( $\partial_t \rho =0, \partial_t \bu = 0, \bu =0$) of the non-local hydrodynamic equations (Eq. (\ref{NS_FULL})).
This follows from the equilibrium relation (see Eq. (25b) in Resibois \cite{Resibois1978}) for the hard sphere system
\begin{equation}
\mathbf{J}_{0,eq} = \left( \frac{\rho_{eq} k_B T}{m} \right) \Grad C^{(1)}[\br_1|\rho_{eq}],
\label{Eq:DFT_eq1}
\end{equation}  
where $C^{(1)}$ is the one particle direct correlation function where
\begin{equation}
\ln \rho_{eq} = C^{(1)}[\br_1|\rho_{eq}],
\label{Eq:DFT_eq2}
\end{equation}
see Hansen et al \cite{Hansen2006}. Further if we assume that the local equilibrium relation ${\bf J}_0 = \left( \frac{\rho_{eq} k_B T}{m} \right) \Grad C^{(1)}[\br_1|\rho]$ holds out of equilibrium,
it is easy to see that the Euler order hydrodynamic equations (Eq. \ref{eqn:Euler_eqn}) satisfy: 
\begin{equation}
\frac{\partial}{\partial t} \left\{ \int \rho \bu^2 d \br_1 + \mathcal{F}[\rho] \right\} =0,
\end{equation}
where $\mathcal{F}[\rho]$ is the Helmholtz free energy as a functional of density (see Section \ref{sec:MeanField} for more details).
Thus the dissipative processes at Euler order are such that the total energy is conserved although the kinetic energy is not conserved.


It is worth noting that the stationary solutions depend on the pair distribution function $G_2[\br_1 , \br_2 | \rho]$ alone 
and not on the transport coefficients or the dissipative terms.
The dissipative terms merely change the path to equilibrium and not the equilibrium itself.
Now given a description of the structure of the liquid as a functional of the density field 
the non-local hydrodynamic equations Eq. (\ref{NS_FULL}) can be used to understand the time evolution of the system toward equilibrium.
In fact the equations can be viewed as a time dependent hydrodynamic density functional theory which we call
Kinetic Density Functional Theory (KDFT).

The effect of the nonlocal dissipation terms will be investigated in a future work. Here, we focus on exploring simpler
models that are more accessible numerically to  establish that hydrodynamic models of the type derived here (and also those derived
by  Archer \cite{Archer2009} and Lutsko \cite{Lutsko2012}) are capable
of capturing solid/liquid phase transitions and the flow induced by these transitions. To this end we make the approximation that the only dissipative term is 
 given by the localized tensor
 \begin{equation}
\mathcal{P}^{C}_{ij}  = - \gamma \frac{1}{2} \left( \partial_i u_j+ \partial_j u_i - \frac{2}{3} \delta_{ij} \Div \bu \right) - \kappa \Div \bu
\end{equation}
where $\gamma>0$ is the shear viscosity coefficient and $\kappa$ is the coefficient of bulk viscosity.
In order to further simplify the problem we assume $\kappa = 0$, which implies the compression of the fluid is dissipationless and reversible
 and that dissipation is purely from shear. This gives the additional momentum flux $ \Div \mathcal{P}^C = - \gamma \Delta \bu$ in Eq. (\ref{eqn:Euler_eqn}).
The simplified, compressible non-local hydrodynamic model is now written as :
\begin{equation}
\begin{array}{ll}
\displaystyle\partial _{t}\rho +\nabla _{\mathbf{r_{1}}}\cdot \rho \mathbf{u}
& =0, \\
\displaystyle\partial _{t}(\rho \mathbf{u})+\nabla _{\mathbf{r_{1}}}\cdot
(\rho \mathbf{u} \mathbf{u}) + \Grad \left( \frac{\rho k_B T}{m} \right) & = \displaystyle  \mathbf{J}_{0}+\gamma
\Delta \mathbf{u},%
\end{array}
\end{equation}%

At this point we comment on the validity of the small gradient expansion in the velocity field
that was used to derive the hydrodynamic model.
In the absence of external forces the gradients in the velocity field are zero at equilibrium (in fact $\bu =0$ at equilibrium) while
out of equilibrium the flow field is driven by the gradients in the chemical potential.
These gradients in the chemical potential are small when the system is close to equilibrium indicating
the gradients in the flow field will also be small close to equilibrium even though density gradients are large.
Thus the small gradient expansion (in Eq. \ref{eq_expansion}) in the velocity field used in the derivation is valid when the system is close to equilibrium.
Further, it is observed in the numerical solution of the model (see section \ref{sec:Numerics}) that the 
velocity gradients in the hydrodynamics are in fact much smaller in comparison to the density gradients.
Although one would ideally want to avoid the small gradient expansion in the velocity field the 
approximation in itself is valid for a system close to equilibrium and the 
model derived here is self consistent with the approximation. 



\section{Kinetic Density Functional Theory}
\label{sec:KDFT}

The simplified hydrodynamic equations obtained from the RET
derived in the previous section take the form
\begin{equation}
\begin{array}{ll}
\displaystyle\partial _{t}\rho +\nabla _{\mathbf{r_{1}}}\cdot \rho \mathbf{u}
& =0, \\
\displaystyle\partial _{t}(\rho \mathbf{u})+\nabla _{\mathbf{r_{1}}}\cdot
(\rho \mathbf{u} \mathbf{u}) + \Grad \left( \frac{\rho k_B T}{m} \right) & = \displaystyle  \mathbf{J}_{0}+\gamma
\Delta \mathbf{u},%
\end{array}
\label{KDFT}
\end{equation}%
where $\mathbf{J}_{0}$ is the nonlocal function of the density in Eq. (\ref{J0}).
We now aim to understand the relation of the non-local hydrodynamic model 
with other Dynamic Density Functional Theory approaches.

\subsection{Mean Field Approximated KDFT}
\label{sec:MeanField}
In order to further understand the relation of  KDFT to DDFT we
use the equilibrium relation Eq. (\ref{Eq:DFT_eq2}) to make a local equilibrium approximation 
\begin{equation}
\mathbf{J}_0 \approx \left( \frac{\rho k_B T}{m} \right) \Grad C^{(1)}[\br_1|\rho].
\end{equation}  
The KDFT with this approximation takes the form
\begin{equation}
\begin{array}{ll}
\displaystyle\partial _{t}\rho +\nabla _{\mathbf{r_{1}}}\cdot \rho \mathbf{u}
& =0, \\
\displaystyle\partial _{t}(\rho \mathbf{u})+\nabla _{\mathbf{r_{1}}}\cdot
(\rho \mathbf{u} \mathbf{u})+\nabla _{\mathbf{r_{1}}}\left( \frac{%
\rho k_{B}T}{m}\right) & \displaystyle=\left( \frac{\rho k_{B}T}{m}\right)
\nabla _{\mathbf{r_{1}}}C^{(1)}[\mathbf{r}_{1}|\rho ] + \gamma \Delta \bu.%
\end{array}
\label{Eq:hydrodynamics}
\end{equation}%
The hydrodynamics can now be written in terms of the Helmholtz Free energy
of the system as
\begin{equation}
\begin{array}{rl}
\displaystyle\partial _{t}\rho +\nabla _{\mathbf{r_{1}}}\cdot \rho \mathbf{u}
& =0, \\
\displaystyle\partial _{t}(\rho \mathbf{u})+\nabla _{\mathbf{r_{1}}}\cdot
(\rho \mathbf{u} \mathbf{u}) & =\displaystyle-\frac{\rho }{m}\nabla _{%
\mathbf{r_{1}}}\left( \frac{\delta \mathcal{F}}{\delta \rho }\right) + \gamma \Delta \bu ,%
\end{array}
\label{Eqn:Hydro_damp}
\end{equation}%
where
\begin{equation}
\label{HFE}
\mathcal{F}[\rho ]=\mathcal{F}_{id}[\rho ]+\mathcal{F}_{ex}[\rho ],
\end{equation}
is the Helmholtz free energy as a functional of the density,
\begin{equation}
\mathcal{F}_{id}[\rho ]=k_BT \int d\mathbf{r}\rho (\mathbf{r})(\ln
(\rho (\mathbf{r}))-1)
\end{equation}%
is the ideal gas part of the free energy and $\mathcal{F}_{ex}[\rho]$ is the excess free energy functional (such that $\displaystyle - \frac{1}{k_B T} \frac{\delta \mathcal{F}_{ex}}{\delta \rho} = C^{(1)}$ the one particle direct correlation function \cite{Hansen2006}). This is the underdamped limit of the models derived by Archer \cite{Archer2009} and Lutsko \cite{Lutsko2012}. 

%

The non-local hydrodynamics Eq. (\ref{Eqn:Hydro_damp}) is dissipative with energy
\begin{equation}
\mathcal{E} [\rho, \bu] := \frac{1}{2} \int d\br_1 \rho \mid \bu \mid^2 +  \mathcal{F}[\rho],
\end{equation}
which is  the sum total of the intrinsic Helmholtz free energy of the dense gas 
and the kinetic energy associated with the flow.
In fact the energy is dissipated by the viscous stress with
\begin{equation}
\frac{d \mathcal{E}[\rho,\bu]}{d t} = -\frac{1}{2} \gamma \int \tilde{\mathcal{D}} : \tilde{\mathcal{D}} d \br_1 \leq 0,
\end{equation}
where $\tilde{\mathcal{D}} := \Grad \bu + \Grad^T \bu$.
As noted in the previous section at Euler order this reduces to $\frac{d \mathcal{E}[\rho,\bu]}{d t} = 0$.

Interestingly, in the Stokes limit, the system in Eq. (\ref{Eqn:Hydro_damp}) reduces to a nonlocal, nonlinear partial differential equation for the density:
\begin{equation}
\partial_t\rho + \frac{1}{m\gamma} \Div \left(\rho\Delta_{\br_1}^{-1}\left(\rho\Grad\left(\frac{\delta \mathcal{F}}{\delta \rho }\right)\right)\right)=0,
\label{Eq_stokes}
\end{equation}
and the Helmholtz free energy is dissipated as $\displaystyle{\partial_t \mathcal{F}=-\frac{1}{m\gamma}\int \rho\Grad\frac{\delta \mathcal{F}}{\delta \rho }\cdot
\left(-\Delta_{\br_1}^{-1}\left(\rho \Grad\frac{\delta \mathcal{F}}{\delta \rho }\right)\right)~d\br\le 0}$. It is worth noting that this is different than the
time-dependent density functional theory as derived 
previously in the overdamped limit \cite{Marconi1999, Yoshimori2005, Lutsko2012, Archer2009}.

The energy minimization process associated with the non-local hydrodynamics
allows us to establish that the dynamics approaches an equilibrium state.
It is apparent that the steady state ($\partial
_{t}\rho =0,\partial _{t}\mathbf{u}=0$) density field corresponding to a
stationary velocity field ($\mathbf{u}=0$) is an extremum of the free energy,
for example it satisfies $\frac{\delta \mathcal{F}}{\delta \rho }=0$. 
This is consistent with  CDFT.
Thus the phase transition at equilibrium in a stationary fluid is the same as predicted
by the CDFT. 
However the introduction of the kinetic energy and shear dissipation alters the path 
to equilibrium in comparison to the over-damped dynamics.
It is also easy to see that the approach to equilibrium and the equilibrium state can be altered by driving
the system using an imposed flow or shear.
This makes the simplified Kinetic Density Functional Theory approach of  Eq. (\ref{KDFT}) and Eq. (\ref{Eqn:Hydro_damp})
suitable for studying the effect of flow on freezing.

At this point one may choose any reliable definition of the
excess free energy  and obtain a reasonable theory for
studying the effect of flow on crystallization. Thus we need to 
 estimate the 
pair correlation function $G_{2}$ or equivalently, the direct correlation
function $C^{\left( 1\right) }$.
Such an approximation to $C^{(1)}$ that allows one to develop a theory is determined by the choice of CDFT. 
The simplest form of density functional theory  was first
introduced by Ramakrishnan and Youssouff (RY) \cite{Ramakrishnan1979} and
Haymet and Oxtoby\cite{Haymet1981}.
Further, based on these principles more sophisticated CDFT
models such as the Effective Liquid Approximation of Baus and Colot\cite%
{Colot1985}, the Weighted Density Approximation of Curtin and Ashcroft\cite%
{Curtin1985,Curtin1986}, the Modified Weighted Density Approximation of
Denton and Ashcroft\cite{Denton1989}, Generalized Effective Liquid
Approximation of Baus \cite{Baus1990,Lutsko1990} and Rosenfeld's Fundamental Measure Theory \cite{Rosenfeld1989} have been developed which
provide better quantitative agreement with particle simulations.
 We refer the reader to Lutsko \cite{Lutsko2010} and Lowen et al \cite{Lowen2010} for recent reviews.
However, to maintain simplicity we present an approximation using the Ramakrishnan-Youssouff \cite{Ramakrishnan1979} formalism  to study the dynamics of the simplified KDFT.

%
%
%
%
%
%
%

\subsection{RY-KDFT : Ramakrishnan-Youssef Approximation to KDFT}
\label{sec:RY-KDFT}

 Working with a homogeneous liquid reference state of density $%
\rho _{ref}$, and  expanding $C^{\left( 1\right) }$ about the reference
density we have%
\begin{equation*}
C^{(1)}[\mathbf{r}_{1}|\rho ]=C^{(1)}[\mathbf{r}_{1}|\rho _{ref}]+\int d%
\mathbf{r}_{2}\frac{\delta C^{(1)}[\mathbf{r}_{1}|\rho _{ref}]}{\delta \rho (%
\mathbf{r}_{2})}\delta \rho (\mathbf{r}_{2})+\ldots
\end{equation*}%
Using this expansion and truncating to lowest order
in the expansion we get%
\begin{equation*}
\mathbf{J}_{0}\approx\left( \frac{%
\rho k_{B}T}{m}\right) \nabla _{\mathbf{r_{1}}}C^{(1)}[\mathbf{r}_{1}|\rho ] =\left( \frac{\rho k_{B}T}{m}\right) \int d\mathbf{r}%
_{2}\nabla _{\mathbf{r_{1}}}C^{(2)}[\mathbf{r}_{1},\mathbf{r}_{2}|\rho
_{ref}]\delta \rho (\mathbf{r}_{2})+O\left( \delta \rho^2 \right),
\end{equation*}%
where $C^{(2)} = \frac{\partial C^{(1)}}{\partial \rho}$.
For the hard sphere system an
exact solution for the two particle direct correlation function $C^{(2)}$ is
known for the Percus and Yevick (PY) closure \cite{Hansen2006} for a homogeneous fluid of density $\bar{\rho}$
\begin{equation}
 C^{(2)} (r,\bar{\rho}) =
\left\{ \begin{array}{lc}   c_0 + c_1\left( \frac{r}{\sigma} \right) +c_3\left(  \frac{r}{\sigma} \right)^3 & \qquad 0 \leq r \leq \sigma, \\ \\
   0 & \qquad \text{otherwise}. \\
 \end{array}\right.
 \end{equation}
Here
\begin{equation*}
\begin{array}{ccc}
c_{0}=-\frac{(1+2\eta )^{2}}{(1-\eta )^{4}}, & c_{1}=\frac{6\eta (1+\frac{1}{%
2}\eta )^{2}}{(1-\eta )^{4}}, & c_{3}=\frac{1}{2}\eta c_{0}%
\end{array}%
\end{equation*}%
where $\eta := \frac{\pi}{6}\sigma^3 \rho_{ref}$ is the packing fraction. 
With this estimate of the two particle direct correlation function, the
hydrodynamics given by RY-KDFT becomes Eq. (\ref{KDFT}) with $J_{0}$
approximated by%
\begin{equation*}
\mathbf{J}_{0} \approx \left( \frac{\rho k_{B}T}{m}\right) \int d \br_2 \Grad \left( \Theta( \sigma - r_{12} ) \left(c_o + c_1 \left(\frac{r_{12}}{\sigma} \right) + c_3 \left( \frac{r_{12}}{\sigma}\right)^3 \right) \right) \delta \rho (\br_2 ),
\end{equation*}
when the reference state corresponds to a homogeneous fluid.

A linear stability analysis (see Appendix \ref{APPENDIX_LIN}) of the non-dimensionalized model presented in
Appendix \ref{APPENDIX_NONDIM} shows that the homogeneous fluid
at rest with density $\bar{\rho}$ and $\rho_{ref} = \bar{\rho}$ is linearly stable if 
\begin{equation}
\begin{array}{cc}
(1- \bar{\rho}    \widehat{ C^{(2)} }(k| \bar{\rho})) > 0 & \text{for all $k$},
\end{array}
\end{equation}
where the hat represents the 3-dimensional Fourier transform with Fourier variable $\bar{k}$ ($k= \mid \bar{k} \mid $) and
 $\sigma$ is set to unity by non-dimensionalization (see Appendix \ref{APPENDIX_LIN} for details regarding the expression for $\hat{f}$ and $\widehat{C^{(2)}}$) .
The instability condition has no solution at packing fraction less than one. Hence the homogeneous fluid at rest is always linearly stable.
The same linear stability condition and conclusions were also noted for the over-damped case by Groh and Mulder \cite{Groh1999} 
 for the PY hard sphere liquid (also see Appendix \ref{APPENDIX_LIN}) .
This however does not mean the fluid does not undergo a phase transition.
The hard sphere liquid under the PY approximation is meta-stable and does in fact undergo a phase transition \cite{Dong2006}.

Now following Ramakrishan and Youssef\cite{Ramakrishnan1979} and Haymet and Oxtoby \cite{Haymet1981}, 
we can hypothesize that there exists $\rho _{L}$ and $\rho _{S}$ (real constants such that $\rho_{S}>\rho _{L}$) 
that determine the phase boundaries. Thus
we expect that, as the average density increases, the equilibrium state
transitions from a homogeneous density (liquid) to a co-existence of solid
and liquid phase at $\rho _{L}$ and then to a pure solid phase at $\rho _{S}$.
Then choosing $\rho_{ref} = \rho_L$ one may solve for the phase boundaries.
In this case RY-KDFT reduces to the standard Ramakrishan and Youssef\cite{Ramakrishnan1979} formalism
of CDFT at equilibrium and
phase transition and equilibrium states (solid, liquid or co-existence state) have been characterized to obtain
the corresponding phase diagram for hard spheres (see Dong  et al \cite{Dong2006}).




\section{Numerical Simulation of the RY-KDFT}
\label{sec:Numerics}
In this section we perform numerical simulations that illustrate the ability of the RY-KDFT
to capture the freezing transition of a hard sphere liquid.
For computational simplicity we present simulations in 2-dimensions using $C^{(2)}[k|\rho]$ and $\hat{f}[k|\rho]$ from a 3-dimensional theory using a
3-dimensional Fourier transform.
Since the Fourier transform of a radially symmetric function is also radially symmetric, this process allows us to impose the structure of a 3-dimensional liquid
in our 2-dimensional simulation.
The simulations presented in this section are analogous to the over-damped simulations performed by  Van Teeffelan et al \cite{VanTeeffelen2009},
where hydrodynamic effects were not considered.

We start with a system that is periodic in both x- and y- directions with a homogeneous liquid of packing fraction $\eta$
and initial velocities set to zero.
A nucleate whose average density $\bar{\rho}$ corresponds to the packing fraction $\eta$
is placed in the liquid.
The nucleate is generated by using Gaussians with peaks located on  a triangular lattice with lattice spacing
\begin{equation}
a = ( 2/\sqrt{3} )^{1/2} \bar{\rho}^{1/2}.
\end{equation}
This is done using the following formula for the solid
\begin{equation}
\rho_s(\br) = \Gamma \sum_i \exp\left( -\alpha (\br_1 - R_{i1})^2 + (\br_2 - R_{i2}) )^2 \right),
 \end{equation}
 where ${\bf R}_i = [ R_{i1}, R_{i2}]^T$ lie on a the triangular lattice of spacing $a$,
  $\alpha$ is a constant chosen to be 200, and $\Gamma$ is a constant chosen to ensure that
 the average density is  $\bar{\rho}$. 
The nucleate of size $6\sqrt{3}/2 a \times 6 a$ is placed in a rectangular system of size $6\sqrt{3}/2 a \times 24 a$ with the nucleate surrounded by
homogeneous liquid of density $\bar{\rho}$.
We note that $\rho_{ref}$ is taken to be $\bar{\rho}$ 
and the phenomenological coefficient of viscosity is chosen to be $\gamma =8$.

\begin{figure}[!ht]
\begin{center}
\begin{picture}(450,350)
\put(-30,0){\includegraphics[width = 6.5in]{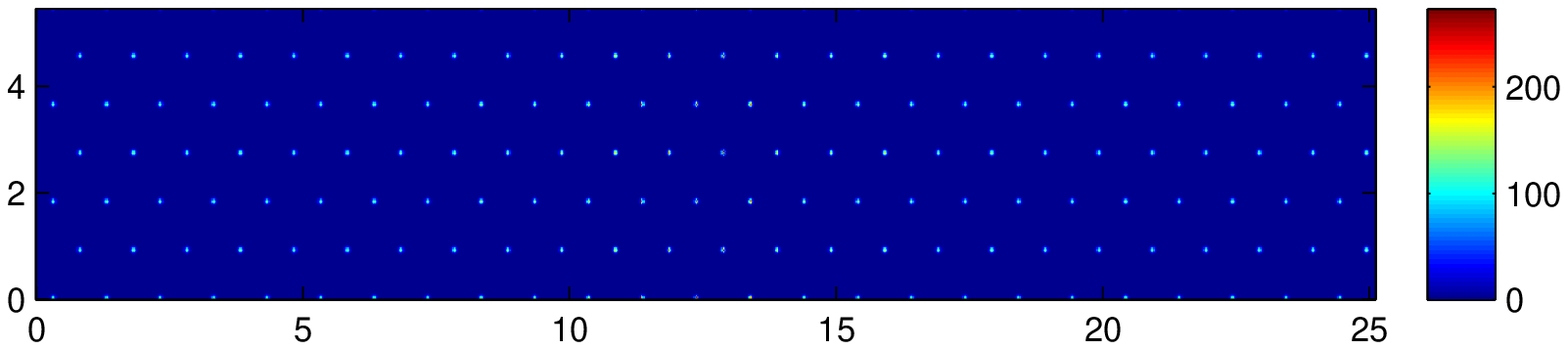}}
\put(-30,115){\includegraphics[width = 6.5in]{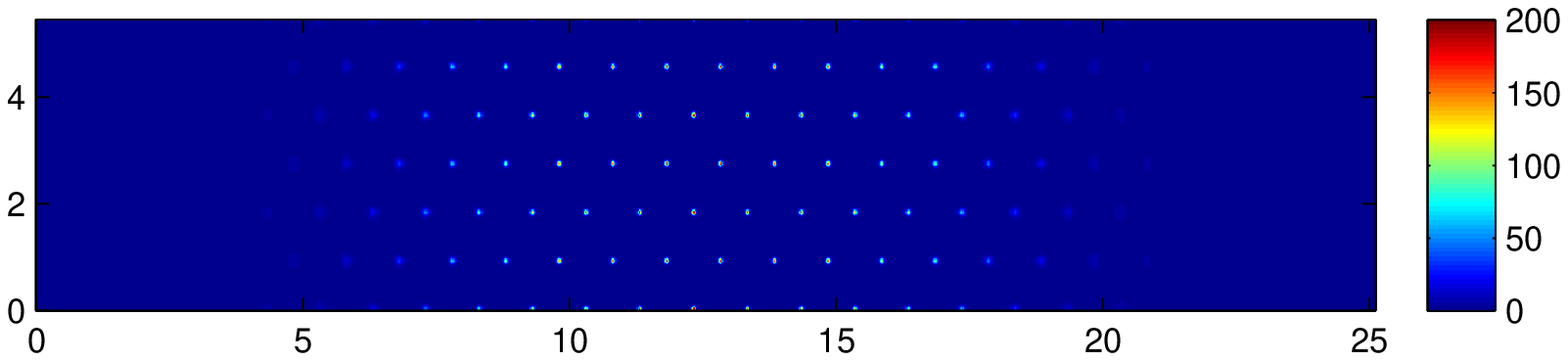}}
\put(-30,235){\includegraphics[width = 6.35in]{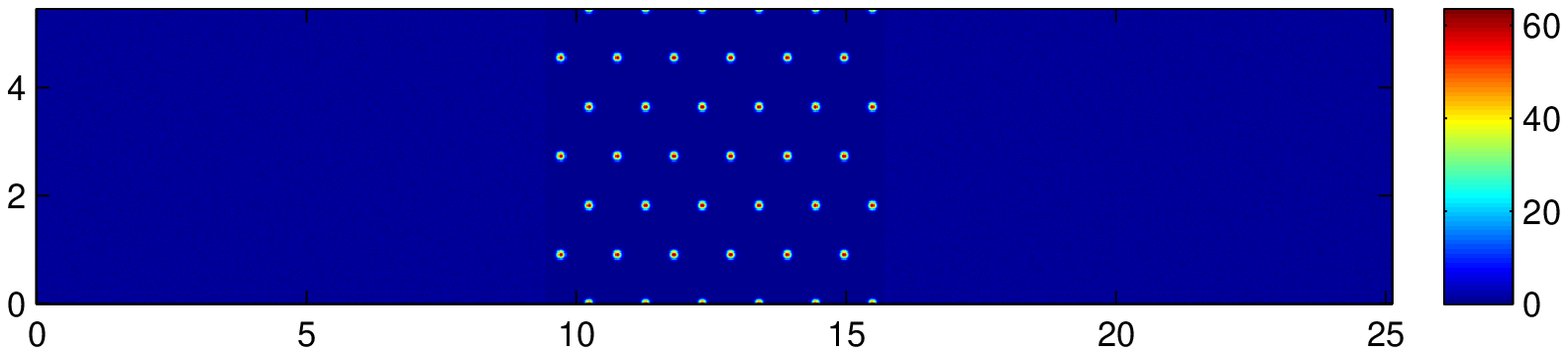}}
\put(430,5){{\bf t=1299.79}}
\put(430,120){{\bf t=291.20}}
\put(430,240){{\bf t=0}}

\end{picture}
\end{center}
\caption{The density field corresponding to growth of a nucleated crystal at packing fraction $\eta = 0.55$ over time using RY-KDFT in section \ref{sec:RY-KDFT}.
The top panel shows the density field at time $t=0$, the middle panel corresponds to $t=291.61$ and the bottom $t=1000.01$. }
\label{fig1}
\end{figure}

\begin{figure}[!ht]
\begin{center}
\begin{picture}(540,480)
\put(-65,0){\includegraphics[width = 3.7in]{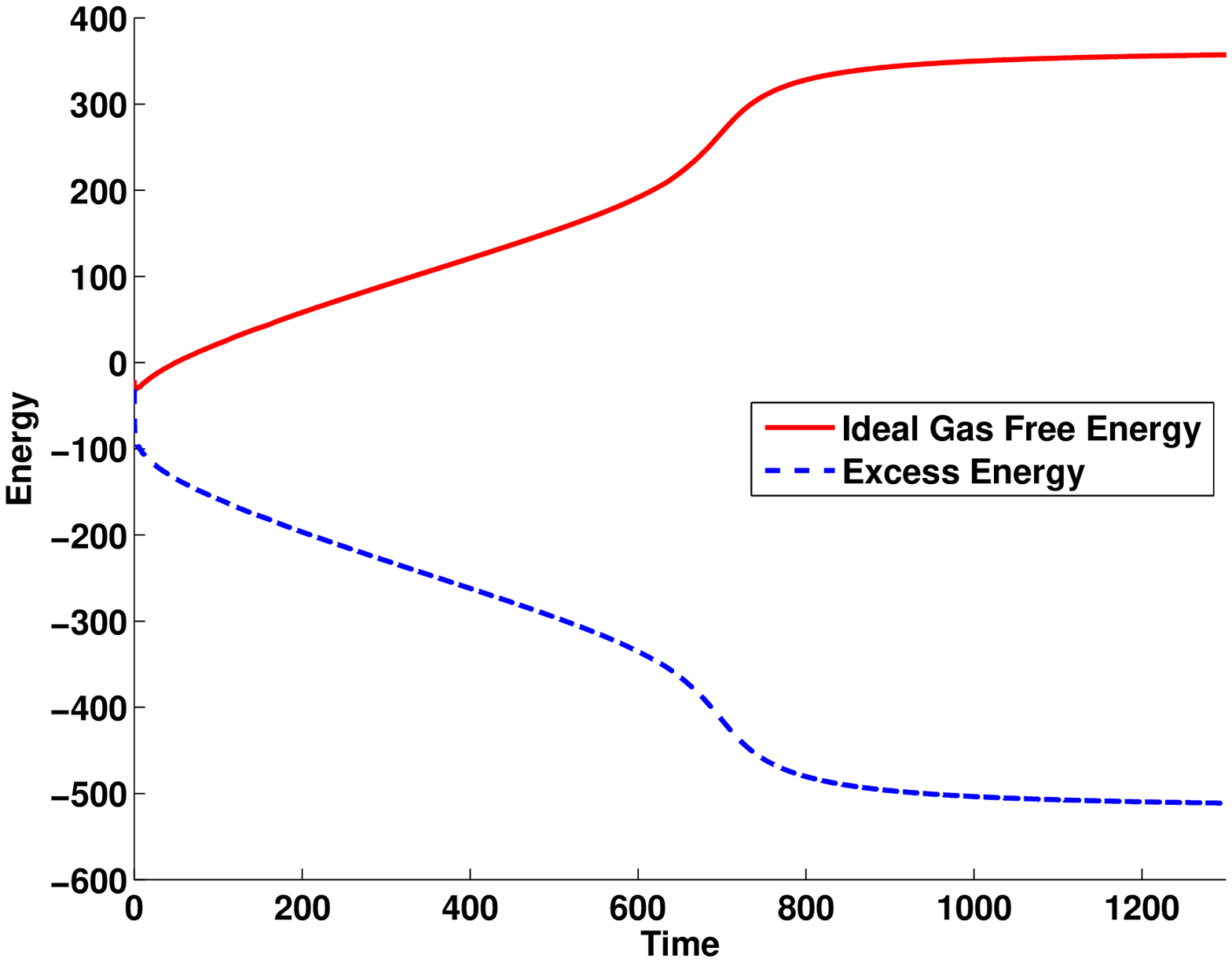}}
\put(195,0){\includegraphics[width = 3.7in]{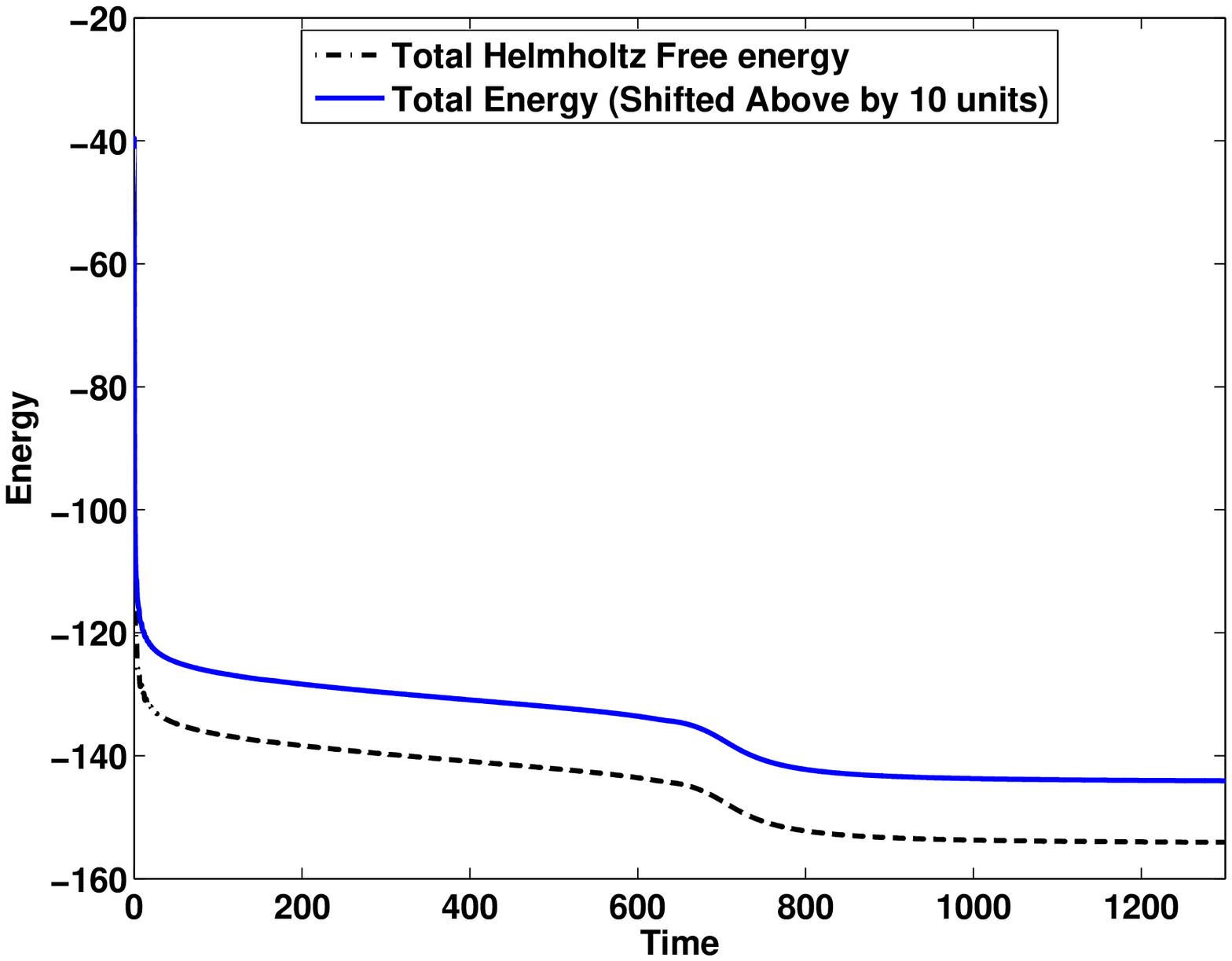}}
\put(60,220){\includegraphics[width = 3.7in]{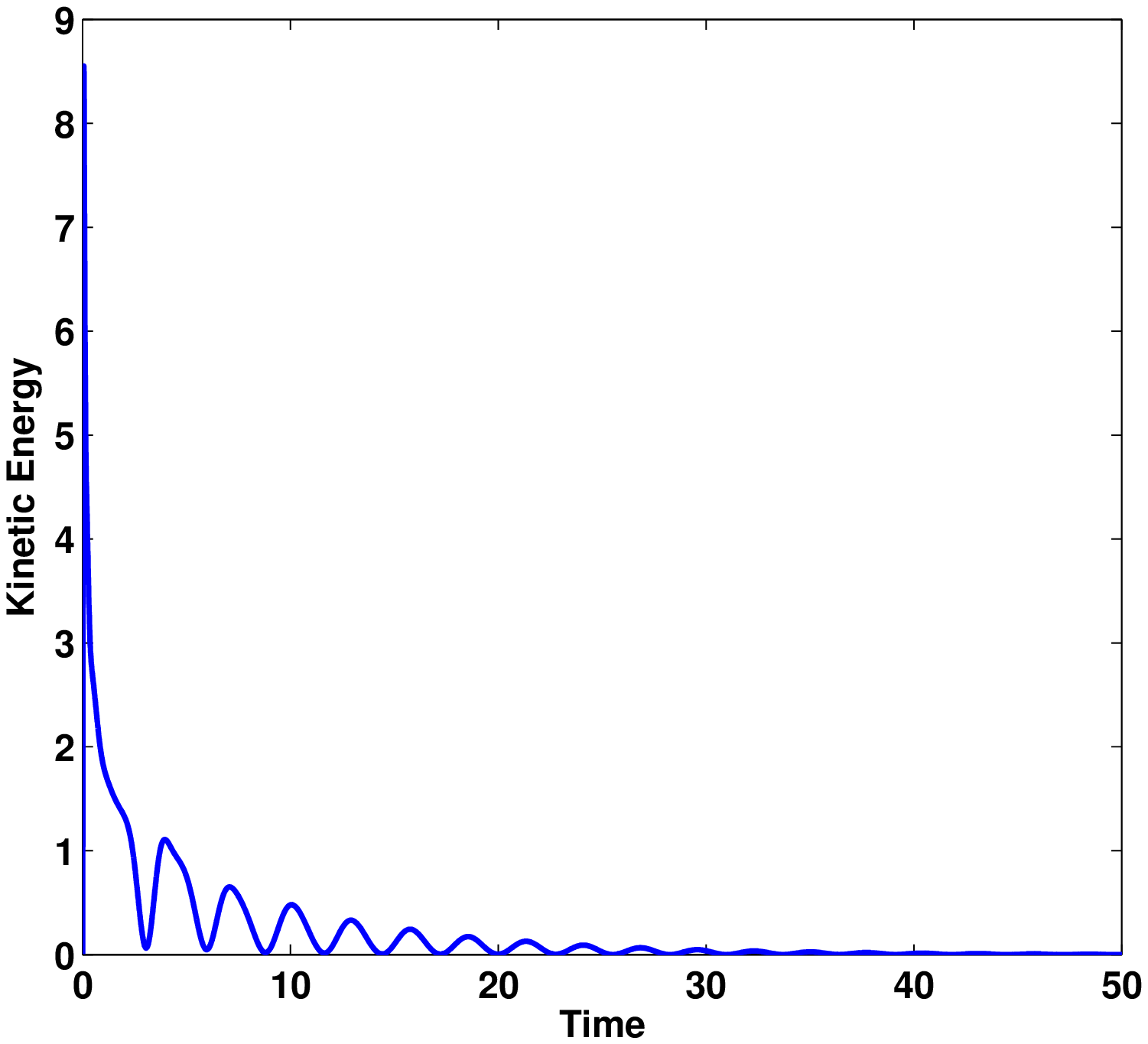}}
\end{picture}
\end{center}
\caption{  The evolution of the total energy (Helmholtz free energy + kinetic energy) and its various components as labelled during the simulation of freezing of RY-KDFT model from Figure \ref{fig1}. The non-monotone evolution of the kinetic energy characterizes the nature of the under damped evolution. }
\label{fig2}

\end{figure}

\begin{figure}[!ht]
\begin{center}
\begin{picture}(400,470)
\put(-50,0){\includegraphics[width = 6.5in]{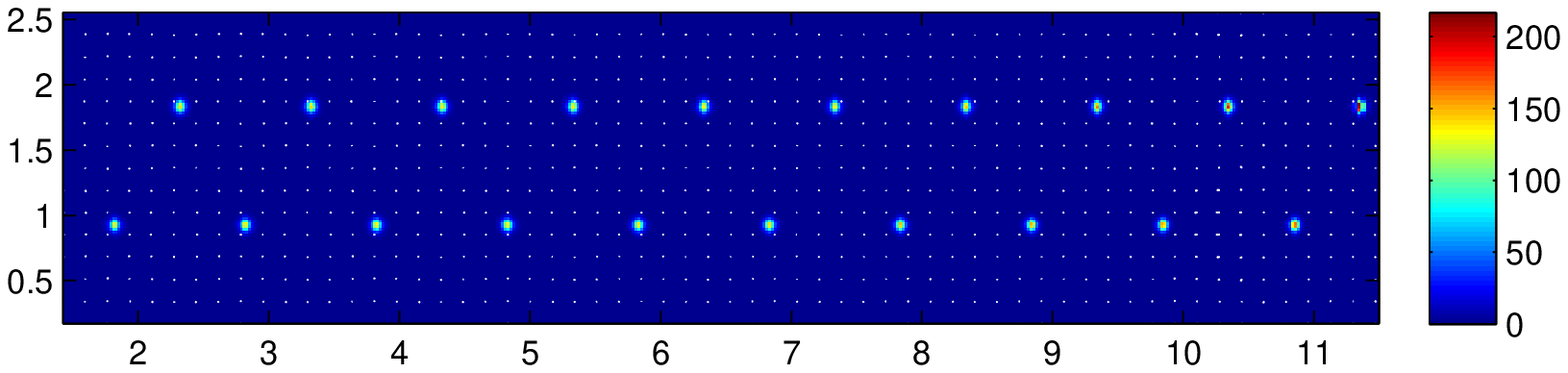}}
\put(-50,120){\includegraphics[width = 6.5in]{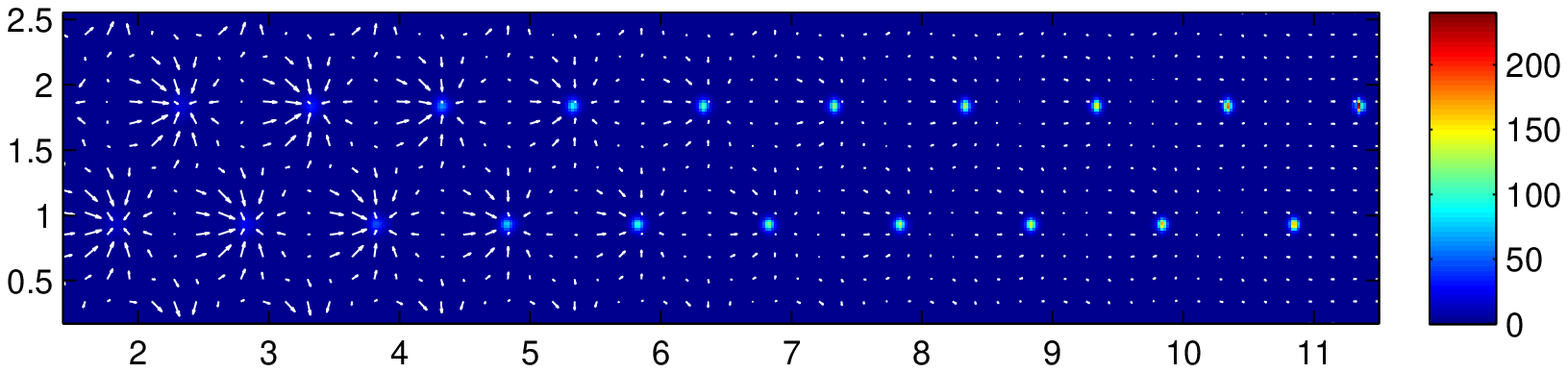}}
\put(-50,240){\includegraphics[width = 6.5in]{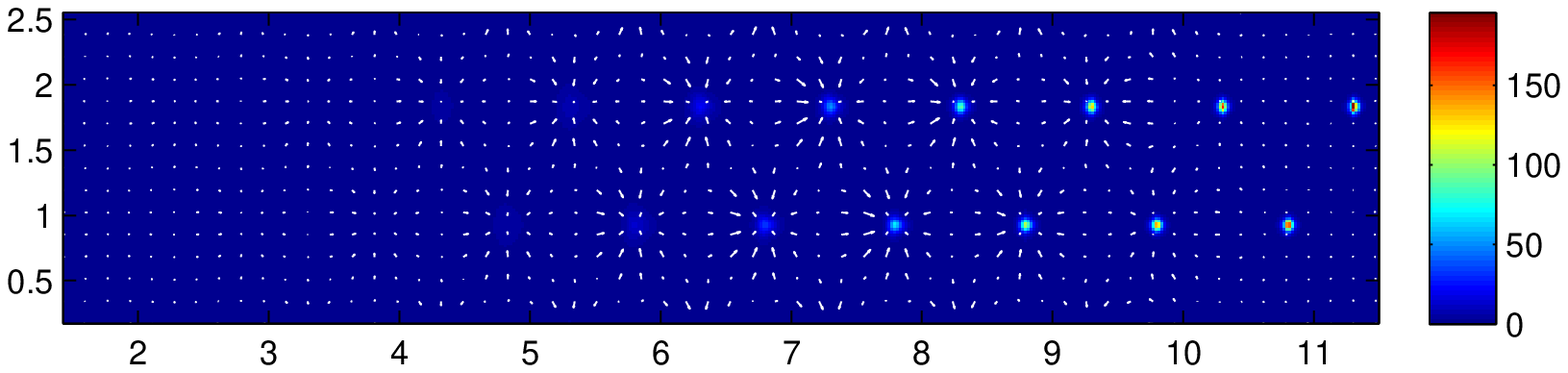}}
\put(-50,360){\includegraphics[width = 6.4in]{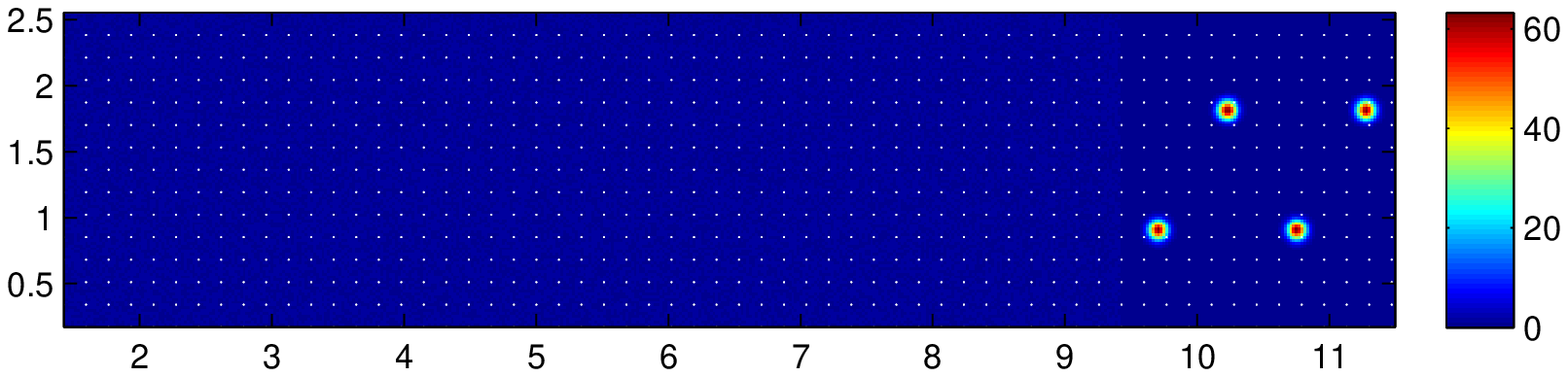}}
\put(410,10){{\bf t = 1299.79}}
\put(410,130){{\bf t=650.35}}
\put(410,250){{\bf t=291.20}}
\put(410,370){{\bf t=0}}
\end{picture}
\end{center}
\caption{A close up of the density field at different times with the corresponding velocity field superimposed for the simulation of the RY-KDFT model reported in Figure \ref{fig1}. The different panels from top to bottom correspond to times $t = 0 ,291.61, 650.53$ and $1000.01$  respectively. The time evolution of the velocity field as the solid liquid interface moves through the regions  shows that the velocity field drives the mass toward the lattice sites where the density field is sharply peaked in the crystal phase. }
\label{fig3}
\end{figure}

\begin{figure}[!ht]
\includegraphics[width = 4.5in]{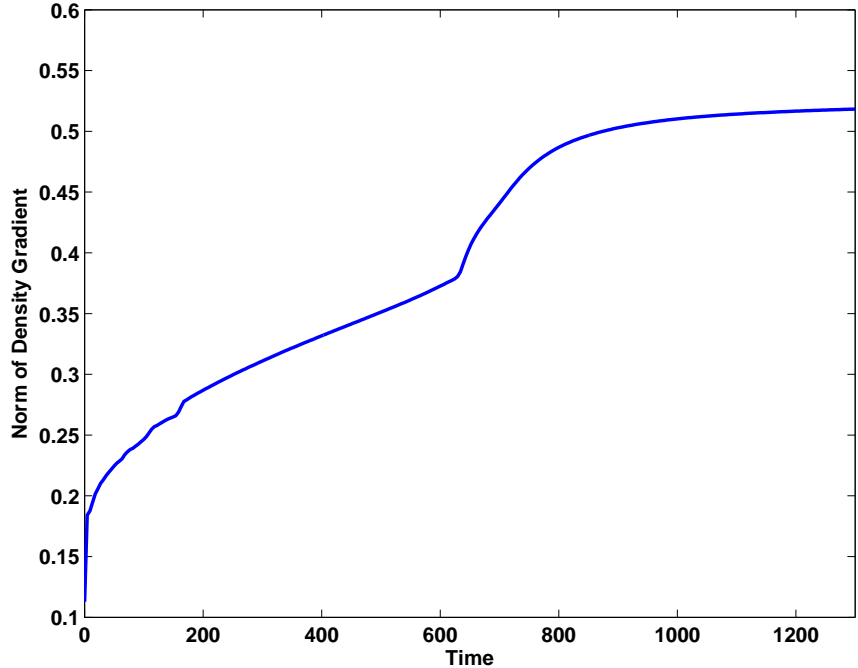}
\caption{The evolution of the $L^2$-norm of the gradient of the density field during the freezing transition in the simulation of RY-KDFT model shown in Figure \ref{fig1}.}
\label{fig4}
\end{figure}

\begin{figure}[!ht]
\includegraphics[width = 4.5in]{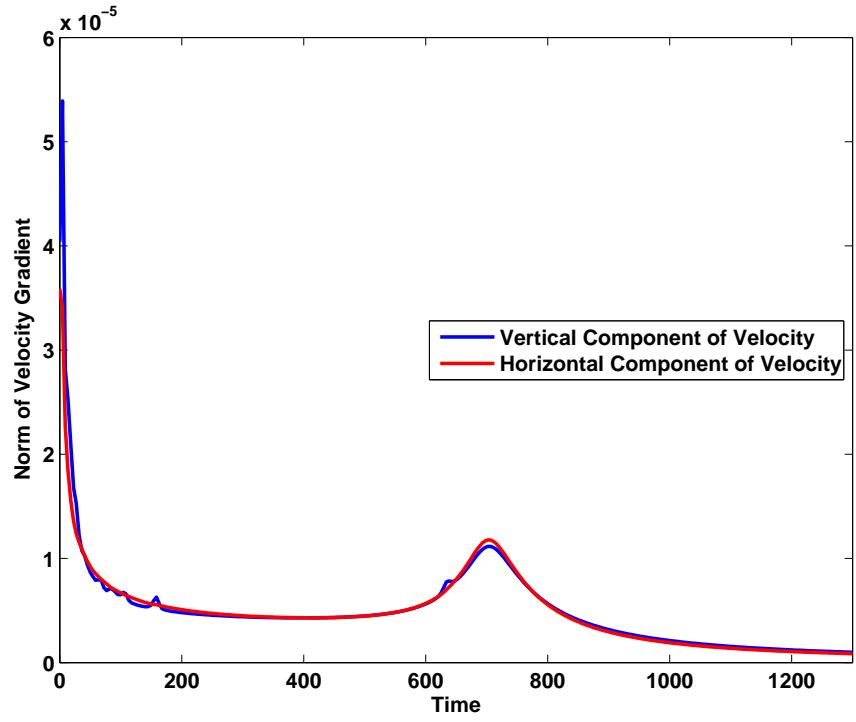}
\caption{The evolution of the $L^2$-norm of the gradient of the velocity field (horizontal and vertical components) during the freezing transition in the simulation of RY-KDFT model shown in Figure \ref{fig1}.}
\label{fig5}
\end{figure}

We now present simulations of RY-KDFT  as the free energy in this case corresponds to the well-studied
CDFT of Ramakrishnan Yousseff \cite{Ramakrishnan1979}. 
Figure \ref{fig1}  shows the time evolution of the RY-KDFT equations at packing fraction $\eta = 0.55$.
The nucleate begins to grow as the system undergoes a liquid to solid phase transition.
The total energy of the system shown in Figure \ref{fig2}  is non-increasing (up to order of numerical accuracy).
However the kinetic energy of the system is not a monotonic function.
The kinetic energy increases periodically while lowering the Helmholtz free energy monotonically.
In particular a closer examination of the Helmholtz free energy shows that the ideal gas part of the free energy
increases while the excess part decreases as one would expect from a freezing transition.
A rapid decrease in the total energy is observed (around $t=600$ to $700$) at the point where the two growing solid liquid interfaces merge to 
produce a complete solid (due to periodic boundary conditions).
This rapid decrease in energy is due to the energetic  advantage to eliminating the solid/liquid interface and
the interfacial energy associated with it. 
This causes a rapid growth in the freezing process seen as a relative increase in the kinetic energy
which is observed in kinetic energy plot in Figure \ref{fig2} and in the velocity field in Figure \ref{fig3} at $t = 650.53$.
Finally the velocity field  of the system varies on the microscopic level
with velocities driving the mass toward the lattice sites where the density is sharply peaked (see Figure. \ref{fig3}).
 While the velocity field  does vary on the scale of the particle it is observed that the gradients in the velocity field
are much smaller than the gradients in the density field (see Figures \ref{fig4} and \ref{fig5}) even out of equilibrium. 
Further the gradients in the density grow with time as expected for the liquid to solid transition but the gradients in the velocity decay
steadily as the system approaches equilibrium.

Our numerical simulations confirm  (not shown) that RY-KDFT predicts the homogeneous liquid is always linearly stable to small perturbations,
but the liquid still undergoes a freezing transition if a nucleate that is large enough is placed in the liquid.
This shows that the homogeneous liquid at packing fraction $\eta =0.55$ is in fact metastable
and that the solid has a lower free energy in comparison to the liquid.




\section{Summary}
A time dependent density functional theory that captures crystal and melt flow
interactions in a dense isothermal gas close to freezing transition has been developed.
Starting with a dense hard gas of interacting particles (hard spheres) the time reversal symmetry of the microscopic equations 
of motion is broken by choosing the Revised Enskog theory as the irreversible equations of motion.
Then using a modified Chapman-Enskog procedure macroscopic equations of motion that
take the form of a non-local hydrodynamic theory is derived which is referred to as Kinetic Density Functional Theory (KDFT).
The relation of the KDFT to classical density functional theory and
 time dependent density functional theories for over-damped systems 
is established.
Based on systematic approximations prescription for a numerically viable theory is presented.
The ability of the model to capture the freezing transition and the flow field associated with the dynamics
is demonstrated through numerical simulations.

\begin{acknowledgements}
Arvind B. and JL gratefully acknowledge partial support from NSF Grants
NSF-CHE 1035218, NSF-DMR 1105409, and NSF-DMS
1217273.
\end{acknowledgements}



\begin{appendix}

\section{Evaluation of the Collisional Contribution to Macroscopic Balance Equations} \label{BALANCE_EQ_APPENDIX}
In this section we derive the collisional contribution to the momentum equation under the ansatz
$\ff( \br_1, \bv_1,t)  = \rho(\br_1,t) \phi (\bv_1 | \bu(\br_1,t))$.
In particular we wish to show that
\begin{equation}
 \int d \bv_1 \bv_1 J_E( G_2[\br_1, \br_2| \rho],f(1), f(2) )  = {\bf J} + \Div \mathcal{P}^C .
\end{equation}
Algebraic manipulations show that 
\begin{equation}
 \int d \bv_1 \bv_1 J_E( G_2[\br_1, \br_2| \rho],f(1), f(2) )  = \mathcal{M}_1 + \mathcal{M}_2
\end{equation}
where 
\begin{equation}
\label{Eqn:M1}
\begin{array}{rl}
\mathcal{M}_1 & = \displaystyle \int d%
\bv_1d \mathbf{v}_{2}d\hat{\bsigma}\sigma ^{2} \bv_1 \Theta (\hat{\mathbf{g}}\bcdot\hat{%
\sigma}) ( \mathbf{g}\bcdot\hat{\bsigma})  \rho(\br_1) \\
&\displaystyle \qquad \qquad \left(G_{2}[\mathbf{r}_1,\mathbf{r}_1-\bsigma \mid \rho (t)] \rho( \br_1 -\bsigma) \phi(\br_1, \bv_1') \phi(\br_1, \bv_2')  \right.
\\
&\displaystyle \qquad \qquad \qquad \left. -G_{2}[\mathbf{r}_1,\mathbf{r}_1+\bsigma \mid \rho (t)] \rho( \br_1 +\bsigma)\phi(\br_1, \bv_1) \phi(\br_1, \bv_2)  \right) . 
\end{array}
\end{equation}
\begin{equation}
\label{Eqn:M2}
\mathcal{M}_2  =
 \int d\bv_1 d \bv_2 d \hat{\bsigma} \sigma^2 \bv_1\Theta ( \hat{\bsigma} \cdot \bg) ( \bg \cdot \hat{\bsigma} )
( F( \br_1, \bv_1' , \br_1 - \bsigma, \bv_2') - F( \br_1, \bv_1, \br_1+ \bsigma, \bv_2) ),
\end{equation}
where $F$ is defined in Eq. (\ref{eq_def_F}).
Now we will show that $\mathcal{M}_1 = {\bf J}$ and $\mathcal{M}_2 = \Div \mathcal{P}^C $.

First let us consider $\mathcal{M}_1$ defined in Eq. (\ref{Eqn:M1}).
$(\bv_1, \bv_2) \to  (\bv_2, \bv_1)$ in the integral (see Eq. (\ref{Eqn:M1})) and adding it to Eq. (\ref{Eqn:M1}) we have
\begin{equation}
\begin{array}{rl}
\mathcal{M}_1 & = \displaystyle\frac{1}{2} \int d%
\bv_1d \mathbf{v}_{2}d\hat{\bsigma}\sigma ^{2}\bg \Theta (\hat{\mathbf{g}}\bcdot\hat{%
\sigma}) ( \mathbf{g}\bcdot\hat{\bsigma}) \rho(\br_1) \\
&\displaystyle \qquad \qquad \left(G_{2}[\mathbf{r}_1,\mathbf{r}_1-\bsigma \mid \rho (t)] \rho( \br_1 -\bsigma) \phi(\br_1, \bv_1') \phi(\br_1, \bv_2')  \right.
\\
&\displaystyle \qquad \qquad \qquad \left. -G_{2}[\mathbf{r}_1,\mathbf{r}_1+\bsigma \mid \rho (t)] \rho( \br_1 +\bsigma)\phi(\br_1, \bv_1) \phi(\br_1, \bv_2)  \right) . 
\end{array}
\end{equation}
By relabeling $(\bv_1',\bv_2') \to (\bv_1,\bv_2)$ and then using the change of variables $(\bv_1',\bv_2') \to (\bv_1,\bv_2)$ in the first
part of the integral,
we get 
\begin{equation}
\begin{array}{rl}
\mathcal{M}_1 & =  \displaystyle \frac{1}{2} \int d%
\bv_1d \mathbf{v}_{2}d\hat{\bsigma}\sigma ^{2}(\bg - 2 \hat{\bsigma}  ( \hat{\bsigma} \cdot \bg)  )  \Theta (-\hat{\bsigma} \cdot \bg) ( \hat{\bsigma} \cdot \bg) \rho(\br_1) \\
&\qquad \qquad \qquad \qquad G_{2}[\mathbf{r}_1,\mathbf{r}_1-\bsigma \mid \rho (t)] \rho( \br_1 -\bsigma) \phi(\br_1, \bv_1) \phi(\br_1, \bv_2)  
\\
&\qquad -  \displaystyle \frac{1}{2} \int d%
\bv_1d \mathbf{v}_{2}d\hat{\bsigma}\sigma ^{2} \bg\Theta (\hat{\mathbf{g}}\bcdot\hat{%
\sigma})( \mathbf{g}\bcdot\hat{\bsigma})  \rho(\br_1) G_{2}[\mathbf{r}_1,\mathbf{r}_1+\bsigma \mid \rho (t)] \rho( \br_1 +\bsigma)\phi(\br_1, \bv_1) \phi(\br_1, \bv_2) , \\
\end{array}
\end{equation}
Next using the change of variables $\hat{\bsigma} \to -\hat{\bsigma}$ in the second integral to simplify the expression,we obtain
\begin{equation}
\begin{array}{rl}
\mathcal{M}_1 & = \displaystyle \int d%
\bv_1d \mathbf{v}_{2}d\hat{\bsigma}\sigma ^{2}  \hat{\bsigma}  ( \hat{\bsigma} \cdot \bg)^2    \Theta (\hat{\bsigma} \cdot \bg)  \rho(\br_1)  G_{2}[\mathbf{r}_1,\mathbf{r}_1-\bsigma \mid \rho (t)] \rho( \br_1 -\bsigma) \phi(\br_1, \bv_1) \phi(\br_1, \bv_2)  \\
& = \displaystyle \frac{1}{2} \int d%
\bv_1d \mathbf{v}_{2}d\hat{\bsigma}\sigma ^{2}  \hat{\bsigma}  ( \hat{\bsigma} \cdot \bg)^2    \rho(\br_1)  G_{2}[\mathbf{r}_1,\mathbf{r}_1-\bsigma \mid \rho (t)] \rho( \br_1 -\bsigma) \phi(\br_1, \bv_1) \phi(\br_1, \bv_2)  \\
& := {\bf J}
\end{array}
\end{equation}
Now we show that $\mathcal{M}_2 =\Div \mathcal{P}^C $.
Using change of variables $(\bv_1, \bv_2) \to  (\bv_2, \bv_1)$ in the integral (see Eq. (\ref{Eqn:M2})) and adding it to Eq. (\ref{Eqn:M2}) we have
\begin{equation}
\label{Manip1}
\begin{array}{ll}
\displaystyle \mathcal{M}_2  & =   \displaystyle \frac{1}{2} \left \{
 \int d\bv_1 d \bv_2 d \hat{\bsigma} \sigma^2 \bv_1 \Theta ( \hat{\bsigma} \cdot \bg) ( \bg \cdot \hat{\bsigma} )
( F( \br_1, \bv_1' , \br_1 - \bsigma, \bv_2') - F( \br_1, \bv_1, \br_1+ \bsigma, \bv_2) )  \right.\\
& \displaystyle \left.  \qquad 
+  \int d\bv_1 d \bv_2 d \hat{\bsigma} \sigma^2 \bv_2 \Theta ( \hat{\bsigma} \cdot \bg) ( \bg \cdot \hat{\bsigma} )
( F( \br_1, \bv_2' , \br_1 - \bsigma, \bv_1') - F( \br_1, \bv_2, \br_1+ \bsigma, \bv_1) ) \right \}.
\end{array}
\end{equation}
Now consider the second integral (call this integral $\mathbb{I}$) and use the change of variables $\hat{\bsigma} \to -\hat{\bsigma}$ to get
\begin{equation}
\label{Manip2}
\begin{array}{rl}
\mathbb{I} & \displaystyle = \int d\bv_1 d \bv_2 d \hat{\bsigma} \sigma^2 \bv_2 \Theta ( \hat{\bsigma} \cdot \bg) ( \bg \cdot \hat{\bsigma} )
( F( \br_1, \bv_2' , \br_1 - \bsigma, \bv_1') - F( \br_1, \bv_2, \br_1+ \bsigma, \bv_1) )  \\
& \displaystyle = - \int d\bv_1 d \bv_2 d \hat{\bsigma} \sigma^2 \bv_2 \Theta ( \hat{\bsigma} \cdot \bg) ( \bg \cdot \hat{\bsigma} )
( F( \br_1, \bv_2' , \br_1 + \bsigma, \bv_1') - F( \br_2-\sigma, \bv_2, \br_2, \bv_1) ).
\end{array}
\end{equation}
Now introducing the variable $\br_{12} = \br_1- \br_2$ we have
\begin{equation}
\begin{array}{rl}
\mathbb{I} & \displaystyle = - \int d\bv_1 d \bv_2 d \hat{\bsigma} d \br_{12} \sigma^2 \bv_2 \Theta ( \hat{\bsigma} \cdot \bg) ( \bg \cdot \hat{\bsigma} )
( \delta(\br_{12} + \bsigma ) F( \br_1, \bv_2' , \br_2 , \bv_1') - \delta(\br_{12} - \bsigma )F( \br_1, \bv_2, \br_2, \bv_1) ) \\
& \displaystyle = - \int d\bv_1 d \bv_2 d \hat{\bsigma} \sigma^2 \bv_2 \Theta ( \hat{\bsigma} \cdot \bg) ( \bg \cdot \hat{\bsigma} )
( F( \br_2-\bsigma, \bv_2' , \br_2, \bv_1') - F( \br_2 + \bsigma, \bv_2, \br_2, \bv_1) ) .
\end{array}
\end{equation}
Finally renaming $\br_1 \to \br_2$ and $\br_2 \to \br_1$ without loss of generality we have
 \begin{equation}
 \label{Manip3}
\mathbb{I} = - \int d\bv_1 d \bv_2 d \hat{\bsigma} \sigma^2 \bv_2 \Theta ( \hat{\bsigma} \cdot \bg) ( \bg \cdot \hat{\bsigma} )
( F( \br_1-\bsigma, \bv_2' , \br_1, \bv_1') - F( \br_1 + \bsigma, \bv_2, \br_1, \bv_1) ) .
\end{equation}
Using the above expression back in Eq. (\ref{Manip1}) we have
\begin{equation}
\mathcal{M}_2 = \frac{1}{2} \int d\bv_1 d \bv_2 d \hat{\bsigma} \sigma^2 \bg \Theta ( \hat{\bsigma} \cdot \bg) ( \bg \cdot \hat{\bsigma} )
( F( \br_1, \bv_1' , \br_1 - \bsigma, \bv_2') - F( \br_1, \bv_1, \br_1+ \bsigma, \bv_2) ) 
\end{equation}
Now relabeling $(\bv_1',\bv_2') \to (\bv_1,\bv_2)$ and then using the change of variables $(\bv_1',\bv_2') \to (\bv_1,\bv_2)$
we get 
\begin{equation}
\begin{array}{ll}
\displaystyle \mathcal{M}_2   &
 \displaystyle = \frac{1}{2} \int d\bv_1 d \bv_2 d \hat{\bsigma} \sigma^2 (\bg - 2 \hat{\bsigma}  ( \hat{\bsigma} \cdot \bg)  )  \Theta (-\hat{\bsigma} \cdot \bg) ( \bg \cdot \hat{\bsigma} )
F( \br_1, \bv_1 , \br_1 - \bsigma, \bv_2)  \\
& \displaystyle  \qquad\qquad - \frac{1}{2} \int d\bv_1 d \bv_2 d \hat{\bsigma} \sigma^2 \bg \Theta ( \hat{\bsigma} \cdot \bg) ( \bg \cdot \hat{\bsigma} )
 F( \br_1, \bv_1, \br_1+ \bsigma, \bv_2)  \\
&  \displaystyle   = - \int d\bv_1 d \bv_2 d \hat{\bsigma} \sigma^2 \hat{\bsigma} \Theta ( \hat{\bsigma} \cdot \bg) ( \bg \cdot \hat{\bsigma} )^2
F( \br_1, \bv_1 , \br_1 + \bsigma, \bv_2)
\end{array}
\end{equation}
where the last step is obtained by changing variables $\hat{\bsigma} \to -\hat{\bsigma}$ in the first integral.
This can further be simplified by  a manipulation similar to Eq. (\ref{Manip2}) and Eq. (\ref{Manip3}) to get
\begin{equation}
\begin{array}{ll}
\displaystyle \mathcal{M}_2 & \displaystyle = - \frac{1}{2} \int d\bv_1 d \bv_2 d \hat{\bsigma} \sigma^2 \hat{\bsigma} \Theta ( \hat{\bsigma} \cdot \bg) ( \bg \cdot \hat{\bsigma} )^2
( F( \br_1, \bv_1 , \br_1 + \bsigma, \bv_2) -F( \br_1 - \bsigma, \bv_1 , \br_1, \bv_2) ) \\
 &\displaystyle = -  \frac{1}{2} \int d\bv_1 d \bv_2 d \hat{\bsigma} \sigma^3 \hat{\bsigma} \Theta ( \hat{\bsigma} \cdot \bg) ( \bg \cdot \hat{\bsigma} )^2
\hat{\bsigma} \bcdot \Grad \int_0^1 d \lambda F( \br_1- (1-\lambda) \bsigma , \bv_1 , \br_1 + \lambda \bsigma, \bv_2) \\
\displaystyle & :=\displaystyle  - \Grad \cdot \mathcal{P}^C

\end{array}
\end{equation}

\section{Chapman-Enskog Method} \label{CE_APPENDIX}

In this appendix we present the details of the expansions involved in the generalized Chapman-Enskog expansion as described in Section \ref{sec:hydrodynamics}.
This expansion assumes the existence of the so called normal solution to the RET 
of the form
\begin{equation}
\label{Eqn:functional}
\ffp{1} = \ff (\bv_1 \mid \rho(\br_1,t), \bu (\br_1,t)).
\end{equation}
The procedure is feasible if the functional dependence in  the velocity $\bu$ can be made local
by expanding the non-local terms, which can be accomplished by assuming 
$ \ff(\bv_1 \mid \rho(\br_1,t), \bu (\br_1,t)) = \rho(\br_1,t) \phi ( \bv_1 | \bu (\br_1,t))$.
We use the following expansions 
\begin{equation}
\begin{array}{rl}
\ff ( \br_1\pm \bsigma, \bv_2,t ) &= \ff (\bv_2 | \rho(\br_1\pm \bsigma,t), \bu ( \br_1\pm \bsigma ,t)) \\
& = \displaystyle \ff (\bv_2 | \rho(\br_1\pm \bsigma,t), \bu ( \br_1 ,t)) \pm \left( \frac{\delta \ff}{\delta \bu} \bcdot  \Grad \bu ( \br_1, \bv_2,t ) \right) \cdot \bsigma + \ldots \\
& = \displaystyle   \ff (\bv_2 | \rho(\br_1\pm \bsigma,t), \bu ( \br_1 ,t))  \\
& \qquad \quad \displaystyle \mp \left( \nabla_{\bv_2} \ff(\bv_2 | \rho(\br_1\pm \bsigma,t),  \bu ( \br_1 ,t)) \bcdot  \Grad \bu ( \br_1, \bv_2,t ) \right) \cdot \bsigma + \ldots ,
\end{array}
\end{equation}
where we have used the relation $\frac{\delta \ff}{\delta \bu} = - \nabla_{\bv_2} \ff(\bv_2 | \rho(\br_1\pm \bsigma,t)$ which  follows from the definition of the variational 
derivative by noting that $\bu = \frac{1}{\rho} \int d \bv_1 \bv_1 \ff ( \br_1\pm \bsigma, \bv_2,t ) $.
This gives 
\begin{equation}
\begin{array}{rl}
\phi ( \bv_1 | \bu ( \br_1, \pm \bsigma,t))
& = \displaystyle \phi (\bv_1 | \bu ( \br_1 ,t)) \pm \left( \frac{\delta \phi}{\delta \bu} \bcdot  \Grad \bu ( \br_1, \bv_1,t ) \right) \cdot \bsigma + \ldots \\
& = \displaystyle   \phi (\bv_1 |  \bu ( \br_1 ,t)) \mp \left( \nabla_{\bv_1} \phi(\bv_2 |  \bu ( \br_1 ,t)) \bcdot  \Grad \bu ( \br_1, \bv_1,t ) \right) \cdot \bsigma + \ldots .
\end{array}
\end{equation}
The gradient expansion of the one particle distribution function above allows us to expand the collision operator as
\begin{equation}
J_E( \ff, \ff_1)  =J_E^0( \ff, \ff_1 ) + J_E^1( \ff, \ff_1 ) + \ldots
\end{equation}
where $f^{(1)} = f^{(1)} ( \br_1,\bv_1,t)$ , $f^{(1)}_1 =   f^{(1)} ( \br_1,\bv_2,t)$,

\begin{equation}
\label{Eqn:localization}
\begin{array}{rll}
J_E^0( \ff, \ff_1)  & =   \displaystyle  \int  & d\bv_2 d\sigh \Theta( \gh \bcdot \sigh ) 
(\bg \bcdot \sigh) \\  
& &\displaystyle \left( G[\br_1,\br_1+\bsigma | \rho] \ff(\bv_1'| \rho (\br_1,t) \bu(\br_1,t) ) \ff(\bv_2'| \rho (\br_1+\bsigma,t) \bu(\br_1,t) ) \right. \\
& & \displaystyle \quad \left. -G[\br_1,\br_1-\bsigma | \rho] \ff(\bv_1| \rho (\br_1,t) \bu(\br_1,t) ) \ff(\bv_2| \rho (\br_1-\bsigma,t) \bu(\br_1,t) )\right) \\
&=   \displaystyle  \int &  d\bv_2 d\sigh \Theta( \gh \bcdot \sigh ) 
(\bg \bcdot \sigh) \\  
& & \displaystyle \left( G[\br_1,\br_1+\bsigma | \rho] \rho (\br_1,t) \phi(\bv_1'|  \bu(\br_1,t) )  \rho (\br_1+\bsigma,t) \phi(\bv_2'| \bu(\br_1,t) ) \right. \\
& & \displaystyle \quad \left. -G[\br_1,\br_1-\bsigma | \rho] \rho (\br_1,t) \phi(\bv_1| \bu(\br_1,t) ) \rho (\br_1-\bsigma,t)  \phi(\bv_2|  \bu(\br_1,t) )\right), 

\end{array}
\end{equation}
and
\begin{equation}
\label{Eqn:JE1}
\begin{array}{rll} J_E^1( \ff, \ff_1 )   = & -   \displaystyle \int  & d\bv_2 d\sigh \Theta( \gh \bcdot \sigh ) (\bg \bcdot \sigh) \\ 
&& \displaystyle \left( G[\br_1,\br_1+\bsigma | \rho] \ff(\bv_1'| \rho (\br_1,t) \bu(\br_1,t) ) \right. \\
&& \quad \left. \left( \nabla_{\bv_2'} \ff_1(\bv_2' | \rho(\br_1+ \bsigma,t) \bu ( \br_1 ,t)) \cdot \Grad \bu( \br_1, \bv_2,t) \right) \cdot \bsigma \right. \\
 & & \displaystyle + \left. G[\br_1,\br_1-\bsigma | \rho] \ff(\bv_1| \rho (\br_1,t) \bu(\br_1,t) ) \right. \\
 && \left. \quad \left( \nabla_{\bv_2} \ff_1(\bv_2 | \rho(\br_1- \bsigma,t) \bu ( \br_1 ,t)) \cdot \Grad \bu( \br_1, \bv_2,t) \right) \cdot \bsigma \right) \\
= & -  \displaystyle \int  & d\bv_2 d\sigh \Theta( \gh \bcdot \sigh ) (\bg \bcdot \sigh) \\ 
 & & \displaystyle \left( G[\br_1,\br_1+\bsigma | \rho] \rho (\br_1,t) \phi(\bv_1'| \bu(\br_1,t) )  \rho(\br_1+ \bsigma,t) \right. \\
 & & \quad \left( \nabla_{\bv_2'} \phi_1(\bv_2' |\bu ( \br_1 ,t)) \cdot \Grad \bu( \br_1, \bv_2,t) \right) \cdot \bsigma \\
 &  & \displaystyle + \left. G[\br_1,\br_1-\bsigma | \rho] \rho (\br_1,t)  \phi(\bv_1|\bu(\br_1,t) )  \rho(\br_1- \bsigma,t) \right. \\
 & & \left.  \left( \nabla_{\bv_2} \phi_1(\bv_2 |\bu ( \br_1 ,t)) \cdot \Grad \bu( \br_1, \bv_2,t) \right) \cdot \bsigma \right).

\end{array} 
\end{equation}

\section{Solution of the Euler Order Distribution} \label{EULER_APPENDIX}
In this appendix we show that the local Maxwellian satisfies the integro-differential equation
\begin{equation}
\left(  ( \bv_1 - \bu ) \cdot \Grad \rho \right) \phi_0 + \left( \bar{\mathcal{P}^{K}}^{(0)}  \cdot \Grad \rho -  {\bf J}_0 \right) \cdot \nabla_{\bv_1} \phi_0 = J_{E}^{\left(
0\right) }(G_{2},f_{0}\left( 1\right) ,f_{0}(2)).
\label{Euler_IE_APP}
\end{equation}
It is easy to see that for the local Maxwellian distribution $\phi^M$ given in Eq.(\ref{eqn:maxwellian}),  $\bar{\mathcal{P}^{K}}^{(0)}_{ij} = \frac{k_BT}{m} \delta_{ij}$  
and $\nabla_{\bv_1} \phi^{M} = -\frac{k_BT}{m} (\bv_1 - \bu) \phi^M$. 
Thus the first and second term in Eq. (\ref{Euler_IE_APP}) cancel reducing the problem to
\begin{equation}
 -  {\bf J}_0  \cdot \nabla_{\bv_1} \phi_0 = J_{E}^{\left(
0\right) }(G_{2},f_{0}\left( 1\right) ,f_{0}(2)).
\label{Euler_IE_reduced}
\end{equation}
By noting that $\mathbf{J}_0 = \int d \bv_1 J^{(0)}_E(G_2, f_0(1), f_0(2))$ and that $f_0 = \rho \phi^M$ one can 
easily perform the velocity integrals leading the expression
\begin{equation}
\mathbf{J}_{0}=\sigma ^{2}\left( \frac{k_{B}T}{m}\right) \rho \left( \mathbf{%
r}_{1}\right) \int d\hat{\bsigma}\hat{\bsigma}\rho \left( \mathbf{r}_{1}-%
\bsigma\right) G_{2}\left[ \mathbf{r}_{1},\mathbf{r}_{1}-\bsigma|\rho \right]
.  \label{J0_APP}
\end{equation}
Now the right hand side of Eq. (\ref{Euler_IE_reduced}) is given by
\begin{equation}
\begin{array}{rl}
J_E^{0} ( \rho \phi^{M}, \rho \phi^{M} )
&  \displaystyle = \int d \bv_2 \int d \hat{\bsigma} \sigma^2  \Theta( \hat{\bsigma} \cdot \bg) ( \hat{\bsigma} \cdot \bg ) \rho( \br_1)\\
& \displaystyle \qquad
\left( G_2[\br_1, \br_1 -\bsigma| \rho] \rho(\br_1-\bsigma)   - G_2[\br_1, \br_1 +\bsigma| \rho] \rho(\br_1+\bsigma) \right)  \phi^M ( \br_1,\bv_1) \phi^{M} ( \br_1, \bv_2), 
\end{array}
\end{equation}
where we have used  the relation $\phi^M(\br_1,\bv_1) \phi^M(\br_1,\bv_2) = \phi^M(\br_1,\bv_1') \phi^M(\br_1,\bv_2')$ to simplify the 
 $J_E^0$ defined in Eq.(\ref{Eqn:localization}). Now changing  variables $-\bsigma \to \bsigma$ in the second half of the integral we obtain
\begin{equation}
\begin{array}{rl}
J_E^{0} ( \rho \phi^{M}, \rho \phi^{M} )
&  \displaystyle = \int d \bv_2 \int d \hat{\bsigma} \sigma^2   ( \hat{\bsigma} \cdot \bg ) \rho( \br_1)\\
& \displaystyle \qquad
G_2[\br_1, \br_1 -\bsigma| \rho] \rho(\br_1-\bsigma) \phi^M ( \br_1,\bv_1) \phi^{M} ( \br_1, \bv_2). 
\end{array}
\end{equation}
Finally  noting that $(\nabla_{\bv_1} - \nabla_{\bv_2}) \phi^M(\br_1,\bv_1) \phi^M(\br_1,\bv_2) = \displaystyle -\frac{k_BT}{m}\bg \phi^M(\br_1,\bv_1) \phi^M(\br_1,\bv_2)$, where $\bg = \bv_1 - \bv_2$
we have
\begin{equation}
\begin{array}{rl}
J_E^{0} ( \rho \phi^{M}, \rho \phi^{M} )
& \displaystyle= - \int d \bv_2 \int d \hat{\bsigma} \sigma^2  \rho( \br_1)  \\
&\displaystyle \qquad
G_2[\br_1, \br_1 -\bsigma| \rho] \rho(\br_1-\bsigma)  \left( \hat{\bsigma} \frac{k_BT}{m}  \cdot \left( \nabla_{\bv_1} - \nabla_{\bv_2} \right)\right) \phi^M ( \br_1,\bv_1) \phi^{M} ( \br_1, \bv_2)
\end{array}
\end{equation}
where the integral over $\bv_2$ can be performed to obtain the equation
\begin{equation}
\begin{array}{rl}
J_E^{0} ( \rho \phi^{M}, \rho \phi^{M} )
&\displaystyle = - \left(  \left(\frac{k_BT}{m}\right)  \rho( \br_1) \int d \hat{\bsigma} \sigma^2 \hat{\bsigma}   
G_2[\br_1, \br_1 -\bsigma| \rho] \rho(\br_1-\bsigma) \right) \cdot \nabla_{\bv_1} \phi^M(\br_1,\bv_1) \\
& \displaystyle  = - \mathbf{J}_0 \cdot \nabla_{\bv_1} \phi^{M} (\br_1,\bv_1).
\end{array}
\end{equation}
Thus we have shown that the local Maxwellian satisfies the integral-differential equation Eq. (\ref{Euler_IE_APP}).



\section{Kinetic Contributions} \label{APPENDIX_KINETIC}
The kinetic contribution to the pressure tensor is given by
\begin{equation}
{\mathcal{P}^{K}}^{(1)} = \int d \bv_1 (\bv_1 - \bu)  (\bv_1 - \bu) \rho \phi_1.
\end{equation}
Since $\phi_1$ is normal to the collisional invariants $1, \bv_1$ and $\mid \bv_1\mid ^2$ we have
\begin{equation}
{\mathcal{P}^{K}}^{(1)}_{ij} = \int d \bv_1 (v_{1i} v_{1j} -\frac{1}{3} \delta_{ij} \mid \bv_1 \mid^2)  \rho \phi_1.
\end{equation}
Now using the form of the solution proposed for $\phi_1$ in Eq. (\ref{CE_FORM}) we have
\begin{equation}
{\mathcal{P}^{K}}^{(1)}_{ij} = \mu^{K}_{ijlm} \mathcal{D}_{lm} + \nu^{K}_{ij} \Div \bu
\end{equation}
where 
\begin{equation}
\label{Eqn:KineticTC1}
\mu^{K}_{ijlm} =\rho   \int d \bv_1 (v_{1i} v_{1j} -\frac{1}{3} \delta_{ij} \mid \bv_1 \mid^2)   \mathcal{C}_{lm} [ \bv_1 \mid \rho]
\end{equation}
and 
\begin{equation}
\label{Eqn:KineticTC2}
\nu^{K}_{ij} = \rho \int d \bv_1 (v_{1i} v_{1j} -\frac{1}{3} \delta_{ij} \mid \bv_1 \mid^2)   \mathcal{Q}[ \bv_1 \mid \rho].
\end{equation}



\section{Collisional Transfer Contributions} \label{APPENDIX_COLLISIONAL_TRANSFER}
The collisional transfer contributions to the pressure tensor are determined from Eq. (\ref{P_collision}).
The pressure tensor at first order in gradients of $\bu$ is given by
\begin{equation}
\begin{array}{rl}
{\mathcal{P}^C} =& \displaystyle \frac{1}{2} \int d \bv_1 d \bv_2 d \hat{\bsigma} \sigma^3 \hat{\bsigma} \hat{\bsigma}  \mid \bg \cdot \hat{\bsigma}\mid^2 \Theta(\bg \cdot \hat{\bsigma})  
 \phi(\br_1,\bv_1) \left[ \left( \nabla_{\bv_2} \phi(\br_1,\bv_2) \cdot \Grad \bu \right) \cdot \bsigma \right] \\
&\displaystyle \qquad  \int_0^1 d \lambda \quad G_2 [ \br_1 -(1-\lambda)\bsigma, \br_1 + \lambda \bsigma ] \rho (\br_1 +\lambda \bsigma) \rho(\br_1 -(1-\lambda)\bsigma) \\
\end{array}
\end{equation}

Now using $\phi = \phi_0 + \phi_1$ and collecting terms to gradient order we get the collisional transfer contributions to the pressure up to 
Navier Stokes order to be
\begin{equation}
\begin{array}{rl}
{\mathcal{P}^C}^{(1)} =& \displaystyle \frac{1}{2} \int d \bv_1 d \bv_2 d \hat{\bsigma} \sigma^3 \hat{\bsigma} \hat{\bsigma}  \mid \bg \cdot \hat{\bsigma}\mid^2 \Theta(\bg \cdot \hat{\bsigma})  \\
&\displaystyle \qquad  \int_0^1 d \lambda \quad G_2 [ \br_1 -(1-\lambda)\bsigma, \br_1 + \lambda \bsigma ] \rho (\br_1 +\lambda \bsigma) \rho(\br_1 -(1-\lambda)\bsigma)  \\
& \displaystyle \qquad \qquad    \phi_0(\br_1,\bv_1) \left[ \nabla_{\bv_2} \phi_0(\br_1,\bv_2) \cdot \Grad \bu \cdot \bsigma \right] .
\end{array}
\label{P_collision_NS}
\end{equation}
Now let us consider the velocity integrals in the above equation :
\begin{equation}
\begin{array}{rl}
\mathcal{M}_3  &:=\displaystyle \int d \bv_1 d \bv_2  \mid \bg \cdot \hat{\bsigma}\mid^2 \Theta(\bg \cdot \hat{\bsigma})  
 \phi_0(\br_1,\bv_1) \nabla_{\bv_2} \phi_0(\br_1,\bv_2) \\
 & =\displaystyle  \left( \frac{-m}{k_B T}\right)\int d \bv_1 d \bv_2  \mid \bg \cdot \hat{\bsigma}\mid^2 \Theta(\bg \cdot \hat{\bsigma})  (\bv_2 -\bu)
  \phi_0(\br_1,\bv_1) \phi_0(\br_1,\bv_2)  \\
 & = \displaystyle \left(\frac{m}{2\pi k_B T}\right)^3 \left( \frac{-m}{k_B T}\right)\int d \bv_1 d \bv_2  \mid \bg \cdot \hat{\bsigma}\mid^2 \Theta(\bg \cdot \hat{\bsigma})  (\bv_2 -\bu)
 \exp\left(-\frac{m(\bv_1 -\bu)^2}{2 k_B T} \right)  \exp(-\frac{m(\bv_2 -\bu)^2}{2 k_B T} )  \\
  & =  \displaystyle \left(\frac{m}{2\pi k_B T}\right)^3 \left( \frac{-m}{k_B T}\right)\int d \bv_1 d \bv_2  \mid \bg \cdot \hat{\bsigma}\mid^2 \Theta(\bg \cdot \hat{\bsigma})  \bv_2  
 \exp\left(-\frac{m\bv_1^2}{2 k_B T} \right)  \exp\left(-\frac{m\bv_2^2}{2 k_B T} \right)  \\
  & =  \displaystyle \left(\frac{m}{2\pi k_B T}\right)^3 \left( \frac{-m}{k_B T}\right)\int d {\bf G} d \bg \mid \bg \cdot \hat{\bsigma}\mid^2 \Theta(\bg \cdot \hat{\bsigma})  \frac{{\bf G} - \bg}{2}  
 \exp\left(-\frac{m{\bf G}^2}{ k_B T} \right)  \exp\left(-\frac{m{\bf g}^2}{4 k_B T} \right) \\
  &= \displaystyle \left(\frac{m}{2\pi k_B T}\right)^3 \left( \frac{m}{k_B T}\right) \left (\int d {\bf G} \exp(-\frac{m{\bf G}^2}{ k_B T} ) \right) \left( \int d \bg \mid \bg \cdot \hat{\bsigma}\mid^2 \Theta(\bg \cdot \hat{\bsigma}) \frac{\bg}{2} \exp( -\frac{ m\bg^2}{4k_B T})  \right) \\
   &= \displaystyle \left(\frac{m}{2\pi k_B T}\right)^3\left( \frac{m}{k_B T}\right) \hat{\bsigma} \left (\int d {\bf G} \exp(-\frac{m{\bf G}^2}{ k_B T} ) \right) \left( \int_0^{\infty} dg g^2 2\pi \int_{-1}^{1} dx \Theta (x) g^2 x^2 \frac{gx}{2} \exp( -\frac{ m g^2}{4 k_B T}) \right)\\
     &= \displaystyle \left(\frac{m}{2\pi k_B T}\right)^3 \left( \frac{m}{k_B T}\right)  \hat{\bsigma}\left (\int d {\bf G} \exp(-\frac{m{\bf G}^2}{ k_B T} ) \right) \left( 2\pi \int_0^{\infty} dg \frac{g^5}{8}\exp( - \frac{m g^2}{4 k_B T} ) \right) \\
     & = \displaystyle8 \left( \frac{m} { \pi k_B T}\right)^{1/2} \hat{\bsigma}
    \end{array}
\end{equation}
where $\bg = \bv_1 -\bv_2$ and ${\bf G} = \frac{\bv_1 +\bv_2}{2}$.
Using the above value for $\mathcal{M}_3$ in Eq. (\ref{P_collision_NS}) we have
\begin{equation}
[\mathcal{P}^C]_{ij} = \mu_{ijkl} \partial_k u_l .
\end{equation}
\begin{equation}
\mu_{ijkl}  = 4 \left( \frac{m} { \pi k_B T}\right)^{1/2} \int  d \hat{\bsigma} \sigma^3 \sigma_i \sigma_j \sigma_k \sigma_l   \int_0^1 d \lambda G_2 [ \br_1 -(1-\lambda)\bsigma, \br_1 + \lambda \bsigma ] \rho (\br_1 +\lambda \bsigma) \rho(\br_1 -(1-\lambda)\bsigma) 
\end{equation}



\section{Calculation of ${\bf J}_1$ Term} \label{APPENDIX_J}  
The non-local contribution from ${\bf J}$ (in Eq. \ref{Eqn:J})  at Navier-Stokes order given by :
\begin{equation}
\begin{array}{ll}
{\bf J}_1& = \displaystyle \frac{1}{2}\int d
\bv_1d \mathbf{v}_{2}d\hat{\bsigma}\sigma ^{2}  \hat{\bsigma}  ( \hat{\bsigma} \cdot \bg)^2     \rho(\br_1)  \rho( \br_1 -\bsigma) G_{2}[\mathbf{r}_1,\mathbf{r}_1-\bsigma \mid \rho (t)]  \\
& \qquad \qquad \qquad \qquad \displaystyle\left[ \phi_0(\br_1, \bv_1) \phi_1(\br_1, \bv_2) +  \phi_1(\br_1, \bv_1) \phi_0(\br_1, \bv_2) \right], \\
& = \displaystyle \int d
\bv_1d \mathbf{v}_{2}d\hat{\bsigma}\sigma ^{2}  \hat{\bsigma}  ( \hat{\bsigma} \cdot \bg)^2    \rho(\br_1)  \rho( \br_1 -\bsigma)G_{2}[\mathbf{r}_1,\mathbf{r}_1-\bsigma \mid \rho (t)] \\
& \qquad \qquad \qquad \qquad \displaystyle\left[ \phi_0(\br_1, \bv_1) \phi_1(\br_1, \bv_2)  \right], \\
& = \displaystyle \int d
\bv_1d \mathbf{v}_{2}d\hat{\bsigma}\sigma ^{2}  \hat{\bsigma}  (  \sigma_i g_i \sigma_j g_j )     \rho(\br_1)\rho( \br_1 -\bsigma)  G_{2}[\mathbf{r}_1,\mathbf{r}_1-\bsigma \mid \rho (t)]  \\
& \qquad \qquad \qquad \qquad \displaystyle\left[ \phi_0(\br_1, \bv_1) \phi_1(\br_1, \bv_2)  \right],
\end{array}
\end{equation}
Using the orthogonality of $\phi_1$ to the collisional invariants $1, \bv_1$ and $\mid \bv_1 \mid^2$ we have
\begin{equation}
\begin{array}{ll}
{\bf J}_1& = \displaystyle  \left( \int d
\bv_1  (v_{1i}v_{1j} - \frac{1}{3} \delta_{ij} \mid \bv_1 \mid^2 )\phi_1(\br_1, \bv_1) \right)  \\
& \qquad \qquad \displaystyle \left(  \int d\hat{\bsigma}\sigma ^{2}  \hat{\bsigma}   \sigma_i \sigma_j  \rho(\br_1) \rho( \br_1 -\bsigma)   G_{2}[\mathbf{r}_1,\mathbf{r}_1-\bsigma \mid \rho (t)] \right).\end{array}
\end{equation}
The above equation can further be simplified as follows :
\begin{equation}
{\bf J}_{1k} = \mathcal{J}_{1ijk}  \left( \int d
\bv_1  (v_{1i}v_{1j} - \frac{1}{3} \delta_{ij} \mid \bv_1 \mid^2 )\phi_1(\br_1, \bv_1) \right) 
\end{equation}
where 
\begin{equation}
\mathcal{J}_{1ijk} = \int d\hat{\bsigma}\sigma ^{2} \sigma_k   \sigma_i \sigma_j  \rho(\br_1)\rho( \br_1 -\bsigma)   G_{2}[\mathbf{r}_1,\mathbf{r}_1-\bsigma \mid \rho (t)].
\end{equation}

Now using the form of the solution proposed in Eq. (\ref{CE_FORM}) we obtain
 
\begin{equation}
\begin{array}{ll}
{\bf J}_{1k}
&=  \mathcal{J}_{1ijk}  \rho \left( \int d
\bv_1 d\hat{\bsigma} (v_{1i}v_{1j} - \frac{1}{3} \delta_{ij} \mid \bv_1 \mid^2 ) ( \mathcal{C}_{lm}[\bv_1] \mathcal{D}_{lm} + \mathcal{Q}[\bv_1] \Div \bu \right) \\
& =    \mathcal{J}_{1ijk} \left( \mu^K_{ijlm} \mathcal{D}_{lm} + \nu^K_{ij} \Div \bu \right) 
\end{array}
\end{equation}
where $\mu^K_{ijlm}$ and $\nu^{K}_{ij}$ are the kinetic contributions to the transport coefficients
defined in Appendix \ref{APPENDIX_KINETIC} in Eq. \ref{Eqn:KineticTC1} and Eq. \ref{Eqn:KineticTC2}.



\section{Calculation of $J_E^{(1)}$ Term} \label{APPENDIX_JE}  
The contribution of the collision operator $J_E$ to the Kinetic Theory at Navier Stokes order is given by (see Appendix \ref{CE_APPENDIX} Eq. (\ref{Eqn:JE1}))
\begin{equation}
\begin{array}{rl}
J_E^{(1)}[ \rho \phi_0, \rho \phi_0] 
= & -   \displaystyle \int  d\bv_2 d\sigh \Theta( \gh \bcdot \sigh ) (\bg \bcdot \sigh)  \\
& \qquad \qquad \displaystyle \left( G[\br_1,\br_1+\bsigma | \rho] \rho (\br_1,t)  \rho(\br_1+ \bsigma,t) \phi_0(\br_1,\bv_1' )  \right. \\
& \qquad \qquad\quad \left. \left( \nabla_{\bv_2'} \phi_0(\br_1,\bv_2')  \cdot \Grad \bu( \br_1,t) \right) \cdot \bsigma \right) \\
 &\qquad \qquad \displaystyle + \left. G[\br_1,\br_1-\bsigma | \rho] \rho (\br_1,t) \rho(\br_1- \bsigma,t) \phi_0(\br_1,\bv_1)   \right. \\
 &\qquad \qquad \left.  \left( \nabla_{\bv_2} \phi_0(\br_1,\bv_2) \cdot \Grad \bu( \br_1,t) \right) \cdot \bsigma \right), \\
 &= \mathcal{K}[ \bv_1\mid \rho] : \Grad \bu (\br_1,t),
\end{array}
\end{equation}
where 
\begin{equation}
\begin{array}{rl}
\mathcal{K}_{ij}[ \bv_1 \mid \rho ] =  &  \displaystyle \int  d\bv_2 d\sigh \Theta( \gh \bcdot \sigh ) (\bg \bcdot \sigh)  \\
& \displaystyle \left( G[\br_1,\br_1+\bsigma | \rho] \rho (\br_1,t) \phi_0(\br_1,\bv_1 ')  \phi_0(\br_1,\bv_2' ) \rho(\br_1+ \bsigma,t)  \frac{(\bv_2' - \bu)_i}{k_BT}  \sigma_j  \right. \\
 & \displaystyle + \left. G[\br_1,\br_1-\bsigma | \rho] \rho (\br_1,t)  \phi_0(\br_1,\bv_1)  \phi_0(\br_1,\bv_2 )  \rho(\br_1- \bsigma,t)    \frac{(\bv_2 - \bu)_i}{k_BT} \sigma_j \right).
 \end{array}
\end{equation}


\section{Non-dimensionalization of the Hydrodynamic Equations}
\label{APPENDIX_NONDIM}
We take a characteristic temperature scale $\mathcal{T}$ to define a characteristic energy $\epsilon := k_B \mathcal{T}$.
The characteristic length scale is chosen to be the particle size $\sigma$.
This defines a characteristic velocity $U = \sqrt{\frac{\epsilon}{m}}$ and a characteristic time $\sigma/U$.
Now we introduce the following non-dimensional variables

\begin{eqnarray}
\rho' = \rho \sigma^3, \\
\bu' = \frac{\bu}{\sqrt{\epsilon/m}}, \\
\br' = \frac{\br}{\sigma}, \\
t' = \frac{t}{\sigma \sqrt{m/\epsilon}}, \\
k_BT' = \frac{k_B T}{\epsilon}, \\
\nu' = \frac{\nu}{\epsilon/\sigma^3}, \\
\gamma' = \gamma \sqrt{\frac{\epsilon}{m}}\frac{1}{\sigma^2}, \\
(C^{2})' = C^{2} \sigma^3, \text{and} \\
(g[1|\rho_{ref}] f)' = g[1|\rho_{ref}] f \sigma^3 .
\end{eqnarray}
The non-dimensionalized equations are written below, where we have dropped the primes for 
ease of exposition. 
\begin{equation}
\label{eq:non_dim}
\begin{array}{rl}
\displaystyle \DPT \rho + \Div \rho \bu &= 0 \\
\displaystyle \rho \left( \DPT \bu + \bu \cdot \Grad \bu +  \nu  \bu \right) + \Grad (k_BT \rho) & =  \displaystyle k_B T \rho \Grad (C * \rho) + \gamma \Delta \bu
 \end{array} 
\end{equation}
Here the momentum equation has been rewritten with some simple manipulations 
and  $F*G$ represents the spatial convolution of the functions $F$ and $G$.
Here ${C} = C^{(2)}[r_{12} | \rho_{ref}]$ for the RY-KDFT.



\section{Linear Stability Analysis}
\label{APPENDIX_LIN}
This appendix details the linear stability analysis of the hydrodynamic equations Eq. (\ref{eq:non_dim})
and its over-damped limit.
We consider a locally perturbed constant density field $\rho = \bar{\rho} + \tilde{\rho}$ and $\bu = 0 + \tilde{ \bu}$.
Here $\tilde{ \rho}$ and  $\tilde{\bu}$ are assumed to be a small local perturbation and $\bar{\rho}$ is the spatial average of $\rho$
and the spatial average of $\tilde{\bu}$  vanishes.
Substituting these into the Eq. (\ref{eq:non_dim}) and collecting terms that are linear
 in $\delta \rho, \delta \bu$ and their derivatives we obtain the linearized dynamics :  
 \begin{equation}
 \label{Eq:Lin_Hyd}
 \begin{array}{c}
 \partial_t \tilde{ \rho} + \rho_0 \Div (\tilde{ \bu}) = 0 \\
 \partial_t (\tilde{ \bu} ) +\nu \tilde{ \bu} = - \frac{k_BT}{\rho_0} \Grad( \tilde{ \rho} ) +k_BT \Grad (C * \tilde{ \rho}) + \frac{\gamma}{\bar{\rho}} \Grad^2 \tilde{ \bu} .
 \end{array}
 \end{equation}
  The Fourier transform (in space) of this linear system of differential equations is
 \begin{equation}
 \left[ \begin{array}{c}
 \displaystyle \partial_t \widehat{\delta \rho} \\ \\
\displaystyle \partial_t \widehat{\delta u_1} \\ \\
\displaystyle \partial_t \widehat{ \delta u_2} \\  \\
\displaystyle\partial_t \widehat{\delta u_2} \end{array} \right]
 = 
\left[ \begin{array}{cccc} 0 & i k_1 \rho_0 & ik_2 \rho_0 & ik_3 \rho_0 \\ \\ 
				     \displaystyle -\frac{ik_1}{\rho_0} k_B T ( 1- \rho_0 \widehat{C} ) & -\nu - \frac{\gamma k^2}{\bar{\rho}} & 0 &0 \\\\
				      \displaystyle-\frac{ik_2}{\rho_0} k_B T ( 1- \rho_0 \widehat{C} ) & 0 & -\nu - \frac{\gamma k^2}{\bar{\rho}}&0 \\ \\
				      \displaystyle-\frac{ik_3}{\rho_0} k_B T ( 1- \rho_0 \widehat{C} ) & 0 &0 & -\nu- \frac{\gamma k^2}{\bar{\rho}} \end{array} \right] 
 \left[ \begin{array}{c}
  \displaystyle \widehat{\delta \rho} \\ \\
 \displaystyle \widehat{\delta u_1} \\  \\
 \displaystyle \widehat{\delta u_2} \\  \\
 \displaystyle \widehat{\delta u_2} \end{array} \right]				      
  \end{equation}
where $\vec{k}= ( k_1,k_2,k_3)^T$ is the Fourier variable and $k = \mid \vec{k} \mid$, the hat represents the Fourier transform 
$\hat{h} := \int_{\mathbb{R}^3} h e^{-i\vec{k} \cdot {\bf{x}}} d \bf{x} $
and $\delta u_i$ represent the components of $\delta \bu$.
The eigenvalues of the above matrix are 
\begin{equation}
\begin{array}{c}
\lambda_1 = \lambda_2 = - \frac{(\rho_0 \nu + \gamma k^2)}{\rho_0} \\
\lambda_3, \lambda_4 =  \frac{ -( \nu \rho_0 + \gamma k^2) \pm \sqrt{-4\rho_0^2k^2 k_B T (1 - \rho_0 \hat{C}) + (\nu\bar{\rho} + \gamma k^2)^2}}{2 \rho_0} 
\end{array}
\end{equation}

Now before we study the stability of the eigenmodes we first note that the constants
$\nu, \bar{\rho}, k_B T \geq 0$.
Since $\nu, \gamma \geq 0$ $\lambda_1, \lambda_2$ correspond to stable modes.
The real part of $\lambda_4$ is non positive and hence this mode does not contribute to an instability.
However if $( 1- \rho_0 \hat{C})<0$ the $\lambda_3$ eigenvalue corresponds to 
an unstable eigenmode. 
A similar linearization for Eq. (\ref{Eq_overdamp}), the over-damped limit, give us the linearized dynamics

 \begin{equation}
 \partial_t \Delta \rho = \frac{k_B T}{\nu} \Grad^2 ( \Delta \rho - \bar{\rho} C*\Delta \rho ).
 \end{equation}
 The Fourier transform of this equation is
 \begin{equation}
 \partial_t \widehat{\Delta \rho} = - \frac{k_B T}{\nu} k^2 (1 - \bar{\rho} \widehat{ C }) \widehat{\Delta \rho}
 \end{equation}
where $k = \mid \vec{k} \mid$.
The above equation tells us that the modes corresponding to
$1 - \bar{\rho} \widehat{C} < 0$ grow and the others decay.
Thus the structure of the disordered phase is determined by the Fourier transform of $ \widehat{ C}$.
Further the linear stability conditions for the hydrodynamic equations
and the over-damped dynamics are both  $\bar{\rho} \widehat{C}(k| \bar{\rho}) < 1$.
Finally the expressions for $\ \widehat{C}(k| \bar{\rho})$ for RY-KDFT given by
\begin{equation}
\widehat{{C}^{(2)}}(k| \bar{\rho}) =4 \pi \sum_{j=0,1,3} J_j ( k ) I_j( \eta )
\end{equation}
where $\eta = \frac{\pi}{6} \bar{\rho}$, as before and 
\begin{equation}
I_0 ( \eta ) = -(1 +2\eta)^2/(1-\eta)^4, 
\end{equation}
\begin{equation}
I_1 ( \eta ) = 6\eta(1+\frac{1}{2}\eta^2)^2/(1-\eta)^4, 
\end{equation}
\begin{equation}
I_3 ( \eta ) = \frac{\eta}{2} I_0 (\eta),
\end{equation}
\begin{equation}
J_0(k) = k^{-3}( \sin(k) - k \cos(k)),
\end{equation}
\begin{equation}
J_1(k) = k^{-4}( (2-k^2) \cos(k)  + 2 k \sin(k) - 2),
\end{equation}
and
\begin{equation}
J_3(k) = k^{-6} ( (12k^2-k^4-24)\cos(k) + (4k^3 -24k) \sin(k) +24 ).
\end{equation}



\end{appendix}


\bibliography{kdft}

\begin{thebibliography}{35}%
\makeatletter
\providecommand \@ifxundefined [1]{%
 \@ifx{#1\undefined}
}%
\providecommand \@ifnum [1]{%
 \ifnum #1\expandafter \@firstoftwo
 \else \expandafter \@secondoftwo
 \fi
}%
\providecommand \@ifx [1]{%
 \ifx #1\expandafter \@firstoftwo
 \else \expandafter \@secondoftwo
 \fi
}%
\providecommand \natexlab [1]{#1}%
\providecommand \enquote  [1]{``#1''}%
\providecommand \bibnamefont  [1]{#1}%
\providecommand \bibfnamefont [1]{#1}%
\providecommand \citenamefont [1]{#1}%
\providecommand \href@noop [0]{\@secondoftwo}%
\providecommand \href [0]{\begingroup \@sanitize@url \@href}%
\providecommand \@href[1]{\@@startlink{#1}\@@href}%
\providecommand \@@href[1]{\endgroup#1\@@endlink}%
\providecommand \@sanitize@url [0]{\catcode `\\12\catcode `\$12\catcode
  `\&12\catcode `\#12\catcode `\^12\catcode `\_12\catcode `\%12\relax}%
\providecommand \@@startlink[1]{}%
\providecommand \@@endlink[0]{}%
\providecommand \url  [0]{\begingroup\@sanitize@url \@url }%
\providecommand \@url [1]{\endgroup\@href {#1}{\urlprefix }}%
\providecommand \urlprefix  [0]{URL }%
\providecommand \Eprint [0]{\href }%
\providecommand \doibase [0]{http://dx.doi.org/}%
\providecommand \selectlanguage [0]{\@gobble}%
\providecommand \bibinfo  [0]{\@secondoftwo}%
\providecommand \bibfield  [0]{\@secondoftwo}%
\providecommand \translation [1]{[#1]}%
\providecommand \BibitemOpen [0]{}%
\providecommand \bibitemStop [0]{}%
\providecommand \bibitemNoStop [0]{.\EOS\space}%
\providecommand \EOS [0]{\spacefactor3000\relax}%
\providecommand \BibitemShut  [1]{\csname bibitem#1\endcsname}%
\let\auto@bib@innerbib\@empty
\bibitem [{\citenamefont {Hansen}\ and\ \citenamefont
  {McDonald}(2006)}]{Hansen2006}%
  \BibitemOpen
  \bibfield  {author} {\bibinfo {author} {\bibfnamefont {J.~P.}\ \bibnamefont
  {Hansen}}\ and\ \bibinfo {author} {\bibfnamefont {I.~R.}\ \bibnamefont
  {McDonald}},\ }\href@noop {} {\emph {\bibinfo {title} {{Theory of Simple
  Fluids}}}},\ \bibinfo {edition} {3rd}\ ed.\ (\bibinfo  {publisher} {Academic
  Press},\ \bibinfo {year} {2006})\BibitemShut {NoStop}%
\bibitem [{\citenamefont {L\"{o}wen}(1994)}]{Lowen1994}%
  \BibitemOpen
  \bibfield  {author} {\bibinfo {author} {\bibfnamefont {H.}~\bibnamefont
  {L\"{o}wen}},\ }\href@noop {} {\bibfield  {journal} {\bibinfo  {journal}
  {Physics Reports}\ }\textbf {\bibinfo {volume} {237}},\ \bibinfo {pages}
  {249} (\bibinfo {year} {1994})}\BibitemShut {NoStop}%
\bibitem [{\citenamefont {Bradsley}\ \emph {et~al.}(1979)\citenamefont
  {Bradsley}, \citenamefont {Hurle},\ and\ \citenamefont
  {Mullin}}]{Bradsley1979}%
  \BibitemOpen
  \bibfield  {author} {\bibinfo {author} {\bibfnamefont {W.}~\bibnamefont
  {Bradsley}}, \bibinfo {author} {\bibfnamefont {J.~D.~T.}\ \bibnamefont
  {Hurle}}, \ and\ \bibinfo {author} {\bibfnamefont {J.~B.}\ \bibnamefont
  {Mullin}},\ }in\ \href@noop {} {\emph {\bibinfo {booktitle} {Crystal Growth :
  A Tutorial Approach}}}\ (\bibinfo  {publisher} {Amsterdam : North-Holland},\
  \bibinfo {year} {1979})\ pp.\ \bibinfo {pages} {157--88}\BibitemShut
  {NoStop}%
\bibitem [{\citenamefont {Hurle}(1977)}]{Hurle1977}%
  \BibitemOpen
  \bibfield  {author} {\bibinfo {author} {\bibfnamefont {D.~T.~J.}\
  \bibnamefont {Hurle}},\ }in\ \href@noop {} {\emph {\bibinfo {booktitle}
  {Crystal Growth and Materials}}},\ \bibinfo {editor} {edited by\ \bibinfo
  {editor} {\bibfnamefont {E.}~\bibnamefont {Kaldis}}\ and\ \bibinfo {editor}
  {\bibfnamefont {H.~J.}\ \bibnamefont {Scheel}}}\ (\bibinfo  {publisher}
  {Amsterdam : North-Holland},\ \bibinfo {year} {1977})\ pp.\ \bibinfo {pages}
  {550--69}\BibitemShut {NoStop}%
\bibitem [{\citenamefont {Solan}\ and\ \citenamefont
  {Ostrach}(1979)}]{Solan1979}%
  \BibitemOpen
  \bibfield  {author} {\bibinfo {author} {\bibfnamefont {A.}~\bibnamefont
  {Solan}}\ and\ \bibinfo {author} {\bibfnamefont {S.}~\bibnamefont
  {Ostrach}},\ }in\ \href@noop {} {\emph {\bibinfo {booktitle} {Preparation and
  Properties of Solid State Material}}},\ \bibinfo {editor} {edited by\
  \bibinfo {editor} {\bibfnamefont {W.~R.}\ \bibnamefont {Wilcox}}}\ (\bibinfo
  {publisher} {New York : Marcel Dekker},\ \bibinfo {year} {1979})\ pp.\
  \bibinfo {pages} {63--110}\BibitemShut {NoStop}%
\bibitem [{\citenamefont {Pimputkar}\ and\ \citenamefont
  {Ostrach}(1981)}]{Pimputkar1981}%
  \BibitemOpen
  \bibfield  {author} {\bibinfo {author} {\bibfnamefont {S.}~\bibnamefont
  {Pimputkar}}\ and\ \bibinfo {author} {\bibfnamefont {S.}~\bibnamefont
  {Ostrach}},\ }\href@noop {} {\bibfield  {journal} {\bibinfo  {journal}
  {Journal of Crystal Growth}\ }\textbf {\bibinfo {volume} {55}},\ \bibinfo
  {pages} {614} (\bibinfo {year} {1981})}\BibitemShut {NoStop}%
\bibitem [{\citenamefont {Glicksman}\ \emph {et~al.}(1986)\citenamefont
  {Glicksman}, \citenamefont {R},\ and\ \citenamefont
  {McFadden}}]{Glicksman1986}%
  \BibitemOpen
  \bibfield  {author} {\bibinfo {author} {\bibfnamefont {M.~E.}\ \bibnamefont
  {Glicksman}}, \bibinfo {author} {\bibfnamefont {C.~S.}\ \bibnamefont {R}}, \
  and\ \bibinfo {author} {\bibfnamefont {G.~B.}\ \bibnamefont {McFadden}},\
  }\href@noop {} {\bibfield  {journal} {\bibinfo  {journal} {Annual Review of
  Fluid Mechanics}\ }\textbf {\bibinfo {volume} {18}},\ \bibinfo {pages} {307}
  (\bibinfo {year} {1986})}\BibitemShut {NoStop}%
\bibitem [{\citenamefont {Ramakrishnan}\ and\ \citenamefont
  {Yussouff}(1979)}]{Ramakrishnan1979}%
  \BibitemOpen
  \bibfield  {author} {\bibinfo {author} {\bibfnamefont {T.}~\bibnamefont
  {Ramakrishnan}}\ and\ \bibinfo {author} {\bibfnamefont {M.}~\bibnamefont
  {Yussouff}},\ }\href@noop {} {\bibfield  {journal} {\bibinfo  {journal}
  {Physical Review B}\ }\textbf {\bibinfo {volume} {19}} (\bibinfo {year}
  {1979})}\BibitemShut {NoStop}%
\bibitem [{\citenamefont {Haymet}\ and\ \citenamefont
  {Oxtoby}(1981)}]{Haymet1981}%
  \BibitemOpen
  \bibfield  {author} {\bibinfo {author} {\bibfnamefont {A.}~\bibnamefont
  {Haymet}}\ and\ \bibinfo {author} {\bibfnamefont {D.}~\bibnamefont
  {Oxtoby}},\ }\href@noop {} {\bibfield  {journal} {\bibinfo  {journal} {The
  Journal of Chemical Physics}\ }\textbf {\bibinfo {volume} {74}},\ \bibinfo
  {pages} {2559} (\bibinfo {year} {1981})}\BibitemShut {NoStop}%
\bibitem [{\citenamefont {Baus}\ and\ \citenamefont {Colot}(1985)}]{Colot1985}%
  \BibitemOpen
  \bibfield  {author} {\bibinfo {author} {\bibfnamefont {M.}~\bibnamefont
  {Baus}}\ and\ \bibinfo {author} {\bibfnamefont {J.}~\bibnamefont {Colot}},\
  }\href@noop {} {\bibfield  {journal} {\bibinfo  {journal} {Molecular
  Physics}\ }\textbf {\bibinfo {volume} {55}},\ \bibinfo {pages} {653}
  (\bibinfo {year} {1985})}\BibitemShut {NoStop}%
\bibitem [{\citenamefont {Curtin}\ and\ \citenamefont
  {Ashcroft}(1985)}]{Curtin1985}%
  \BibitemOpen
  \bibfield  {author} {\bibinfo {author} {\bibfnamefont {W.}~\bibnamefont
  {Curtin}}\ and\ \bibinfo {author} {\bibfnamefont {N.}~\bibnamefont
  {Ashcroft}},\ }\href@noop {} {\bibfield  {journal} {\bibinfo  {journal}
  {Physical Review A}\ }\textbf {\bibinfo {volume} {32}},\ \bibinfo {pages}
  {2909} (\bibinfo {year} {1985})}\BibitemShut {NoStop}%
\bibitem [{\citenamefont {Curtin}\ and\ \citenamefont
  {Ashcroft}(1986)}]{Curtin1986}%
  \BibitemOpen
  \bibfield  {author} {\bibinfo {author} {\bibfnamefont {W.}~\bibnamefont
  {Curtin}}\ and\ \bibinfo {author} {\bibfnamefont {N.}~\bibnamefont
  {Ashcroft}},\ }\href@noop {} {\bibfield  {journal} {\bibinfo  {journal}
  {Physical Review Letters}\ }\textbf {\bibinfo {volume} {56}},\ \bibinfo
  {pages} {2775} (\bibinfo {year} {1986})}\BibitemShut {NoStop}%
\bibitem [{\citenamefont {Denton}\ and\ \citenamefont
  {Ashcroft}(1989)}]{Denton1989}%
  \BibitemOpen
  \bibfield  {author} {\bibinfo {author} {\bibfnamefont {A.}~\bibnamefont
  {Denton}}\ and\ \bibinfo {author} {\bibfnamefont {N.}~\bibnamefont
  {Ashcroft}},\ }\href@noop {} {\bibfield  {journal} {\bibinfo  {journal}
  {Physical Review A}\ }\textbf {\bibinfo {volume} {39}},\ \bibinfo {pages}
  {4701} (\bibinfo {year} {1989})}\BibitemShut {NoStop}%
\bibitem [{\citenamefont {Baus}(1990)}]{Baus1990}%
  \BibitemOpen
  \bibfield  {author} {\bibinfo {author} {\bibfnamefont {M.}~\bibnamefont
  {Baus}},\ }\href@noop {} {\bibfield  {journal} {\bibinfo  {journal} {Journal
  of Physics. Condensed Matter : An Institute of Physics Journal}\ }\textbf
  {\bibinfo {volume} {2}},\ \bibinfo {pages} {SA135} (\bibinfo {year}
  {1990})}\BibitemShut {NoStop}%
\bibitem [{\citenamefont {Lutsko}\ and\ \citenamefont
  {Baus}(1990)}]{Lutsko1990}%
  \BibitemOpen
  \bibfield  {author} {\bibinfo {author} {\bibfnamefont {J.~F.}\ \bibnamefont
  {Lutsko}}\ and\ \bibinfo {author} {\bibfnamefont {M.}~\bibnamefont {Baus}},\
  }\href@noop {} {\bibfield  {journal} {\bibinfo  {journal} {Physical Review
  A}\ }\textbf {\bibinfo {volume} {41}},\ \bibinfo {pages} {6647} (\bibinfo
  {year} {1990})}\BibitemShut {NoStop}%
\bibitem [{\citenamefont {Marconi}\ and\ \citenamefont
  {Tarazona}(1999)}]{Marconi1999}%
  \BibitemOpen
  \bibfield  {author} {\bibinfo {author} {\bibfnamefont {U.}~\bibnamefont
  {Marconi}}\ and\ \bibinfo {author} {\bibfnamefont {P.}~\bibnamefont
  {Tarazona}},\ }\href@noop {} {\bibfield  {journal} {\bibinfo  {journal}
  {Journal of Chemical Physics}\ }\textbf {\bibinfo {volume} {110}},\ \bibinfo
  {pages} {8032} (\bibinfo {year} {1999})}\BibitemShut {NoStop}%
\bibitem [{\citenamefont {Yoshimori}(2005)}]{Yoshimori2005}%
  \BibitemOpen
  \bibfield  {author} {\bibinfo {author} {\bibfnamefont {A.}~\bibnamefont
  {Yoshimori}},\ }\href {\doibase 10.1103/PhysRevE.71.031203} {\bibfield
  {journal} {\bibinfo  {journal} {Physical Review E}\ }\textbf {\bibinfo
  {volume} {71}},\ \bibinfo {pages} {1} (\bibinfo {year} {2005})}\BibitemShut
  {NoStop}%
\bibitem [{\citenamefont {Espa\~{n}ol}\ and\ \citenamefont
  {L\"{o}wen}(2009)}]{Espanol2009}%
  \BibitemOpen
  \bibfield  {author} {\bibinfo {author} {\bibfnamefont {P.}~\bibnamefont
  {Espa\~{n}ol}}\ and\ \bibinfo {author} {\bibfnamefont {H.}~\bibnamefont
  {L\"{o}wen}},\ }\href {\doibase 10.1063/1.3266943} {\bibfield  {journal}
  {\bibinfo  {journal} {The Journal of Chemical Physics}\ }\textbf {\bibinfo
  {volume} {131}},\ \bibinfo {pages} {244101} (\bibinfo {year}
  {2009})}\BibitemShut {NoStop}%
\bibitem [{\citenamefont {Goddard}\ \emph {et~al.}(2012)\citenamefont
  {Goddard}, \citenamefont {Pavliotis},\ and\ \citenamefont
  {Kalliadasis}}]{2012Kaliadasis}%
  \BibitemOpen
  \bibfield  {author} {\bibinfo {author} {\bibfnamefont {B.~D.}\ \bibnamefont
  {Goddard}}, \bibinfo {author} {\bibfnamefont {G.~A.}\ \bibnamefont
  {Pavliotis}}, \ and\ \bibinfo {author} {\bibfnamefont {S.}~\bibnamefont
  {Kalliadasis}},\ }\href@noop {} {\bibfield  {journal} {\bibinfo  {journal}
  {SIAM Multiscale Model. Simul.}\ }\textbf {\bibinfo {volume} {10}},\ \bibinfo
  {pages} {633} (\bibinfo {year} {2012})}\BibitemShut {NoStop}%
\bibitem [{\citenamefont {Archer}(2009)}]{Archer2009}%
  \BibitemOpen
  \bibfield  {author} {\bibinfo {author} {\bibfnamefont {A.~J.}\ \bibnamefont
  {Archer}},\ }\href {\doibase 10.1063/1.3054633} {\bibfield  {journal}
  {\bibinfo  {journal} {The Journal of Chemical Physics}\ }\textbf {\bibinfo
  {volume} {130}},\ \bibinfo {pages} {014509} (\bibinfo {year}
  {2009})}\BibitemShut {NoStop}%
\bibitem [{\citenamefont {Chavanis}(2011)}]{Chavanis}%
  \BibitemOpen
  \bibfield  {author} {\bibinfo {author} {\bibfnamefont {P.-H.}\ \bibnamefont
  {Chavanis}},\ }\href {\doibase http://dx.doi.org/10.1016/j.physa.2010.12.018}
  {\bibfield  {journal} {\bibinfo  {journal} {Physica A: Statistical Mechanics
  and its Applications}\ }\textbf {\bibinfo {volume} {390}},\ \bibinfo {pages}
  {1546 } (\bibinfo {year} {2011})}\BibitemShut {NoStop}%
\bibitem [{\citenamefont {Lutsko}(2012)}]{Lutsko2012}%
  \BibitemOpen
  \bibfield  {author} {\bibinfo {author} {\bibfnamefont {J.~F.}\ \bibnamefont
  {Lutsko}},\ }\href {\doibase 10.1063/1.3677191} {\bibfield  {journal}
  {\bibinfo  {journal} {The Journal of Chemical Physics}\ }\textbf {\bibinfo
  {volume} {136}},\ \bibinfo {pages} {034509} (\bibinfo {year}
  {2012})}\BibitemShut {NoStop}%
\bibitem [{\citenamefont {{Van Beijeren}}\ and\ \citenamefont
  {Ernst}(1973)}]{VanBeijeren1973}%
  \BibitemOpen
  \bibfield  {author} {\bibinfo {author} {\bibfnamefont {H.}~\bibnamefont {{Van
  Beijeren}}}\ and\ \bibinfo {author} {\bibfnamefont {M.}~\bibnamefont
  {Ernst}},\ }\href@noop {} {\bibfield  {journal} {\bibinfo  {journal}
  {Physica}\ }\textbf {\bibinfo {volume} {68}},\ \bibinfo {pages} {437}
  (\bibinfo {year} {1973})}\BibitemShut {NoStop}%
\bibitem [{\citenamefont {Kirkpatrick}\ \emph {et~al.}(1990)\citenamefont
  {Kirkpatrick}, \citenamefont {Das}, \citenamefont {Ernst},\ and\
  \citenamefont {Piasecki}}]{Kirkpatrick1990}%
  \BibitemOpen
  \bibfield  {author} {\bibinfo {author} {\bibfnamefont {T.}~\bibnamefont
  {Kirkpatrick}}, \bibinfo {author} {\bibfnamefont {S.}~\bibnamefont {Das}},
  \bibinfo {author} {\bibfnamefont {M.}~\bibnamefont {Ernst}}, \ and\ \bibinfo
  {author} {\bibfnamefont {J.}~\bibnamefont {Piasecki}},\ }\href@noop {}
  {\bibfield  {journal} {\bibinfo  {journal} {The Journal of Chemical Physics}\
  }\textbf {\bibinfo {volume} {92}},\ \bibinfo {pages} {3768} (\bibinfo {year}
  {1990})}\BibitemShut {NoStop}%
\bibitem [{\citenamefont {Resibois}\ and\ \citenamefont
  {De~Leener}(1965)}]{Resibois1965}%
  \BibitemOpen
  \bibfield  {author} {\bibinfo {author} {\bibfnamefont {P.}~\bibnamefont
  {Resibois}}\ and\ \bibinfo {author} {\bibfnamefont {M.}~\bibnamefont
  {De~Leener}},\ }\href@noop {} {\emph {\bibinfo {title} {{Classical Kinetic
  Theory of Fluids}}}},\ edited by\ \bibinfo {editor} {\bibnamefont
  {McGrawhill}}\ (\bibinfo {year} {1965})\BibitemShut {NoStop}%
\bibitem [{\citenamefont {Lutsko}(1996)}]{Lutsko1996}%
  \BibitemOpen
  \bibfield  {author} {\bibinfo {author} {\bibfnamefont {J.~F.}\ \bibnamefont
  {Lutsko}},\ }\href@noop {} {\bibfield  {journal} {\bibinfo  {journal}
  {Physical Review Letters}\ }\textbf {\bibinfo {volume} {77}},\ \bibinfo
  {pages} {2225} (\bibinfo {year} {1996})}\BibitemShut {NoStop}%
\bibitem [{\citenamefont {Lutsko}(2011)}]{Lutsko2001}%
  \BibitemOpen
  \bibfield  {author} {\bibinfo {author} {\bibfnamefont {J.~F.}\ \bibnamefont
  {Lutsko}},\ }\href@noop {} {\bibfield  {journal} {\bibinfo  {journal}
  {Physical Review E}\ }\textbf {\bibinfo {volume} {63}} (\bibinfo {year}
  {2011})}\BibitemShut {NoStop}%
\bibitem [{\citenamefont {Resibois}(1978)}]{Resibois1978}%
  \BibitemOpen
  \bibfield  {author} {\bibinfo {author} {\bibfnamefont {P.}~\bibnamefont
  {Resibois}},\ }\href@noop {} {\bibfield  {journal} {\bibinfo  {journal}
  {Journal of Statistical Physics}\ }\textbf {\bibinfo {volume} {19}},\
  \bibinfo {pages} {593} (\bibinfo {year} {1978})}\BibitemShut {NoStop}%
\bibitem [{\citenamefont {Piasecki}(1987)}]{Piasecki1987}%
  \BibitemOpen
  \bibfield  {author} {\bibinfo {author} {\bibfnamefont {J.}~\bibnamefont
  {Piasecki}},\ }\href@noop {} {\bibfield  {journal} {\bibinfo  {journal}
  {Journal of Statistical Physics}\ }\textbf {\bibinfo {volume} {48}},\
  \bibinfo {pages} {1203} (\bibinfo {year} {1987})}\BibitemShut {NoStop}%
\bibitem [{\citenamefont {Rosenfeld}(1989)}]{Rosenfeld1989}%
  \BibitemOpen
  \bibfield  {author} {\bibinfo {author} {\bibfnamefont {Y.}~\bibnamefont
  {Rosenfeld}},\ }\href@noop {} {\bibfield  {journal} {\bibinfo  {journal}
  {Phys. Rev. Lett.}\ }\textbf {\bibinfo {volume} {63}},\ \bibinfo {pages}
  {980} (\bibinfo {year} {1989})}\BibitemShut {NoStop}%
\bibitem [{\citenamefont {Lutsko}(2010)}]{Lutsko2010}%
  \BibitemOpen
  \bibfield  {author} {\bibinfo {author} {\bibfnamefont {J.~F.}\ \bibnamefont
  {Lutsko}},\ }\href@noop {} {\bibfield  {journal} {\bibinfo  {journal}
  {Advances in Chemical Physics}\ }\textbf {\bibinfo {volume} {144}},\ \bibinfo
  {pages} {1} (\bibinfo {year} {2010})}\BibitemShut {NoStop}%
\bibitem [{\citenamefont {Haataja}\ \emph {et~al.}(2010)\citenamefont
  {Haataja}, \citenamefont {Gr{\'a}n{\'a}syL{\'a}szl{\'o}},\ and\ \citenamefont
  {L{\"o}wen}}]{Lowen2010}%
  \BibitemOpen
  \bibfield  {author} {\bibinfo {author} {\bibfnamefont {M.}~\bibnamefont
  {Haataja}}, \bibinfo {author} {\bibnamefont {Gr{\'a}n{\'a}syL{\'a}szl{\'o}}},
  \ and\ \bibinfo {author} {\bibfnamefont {H.}~\bibnamefont {L{\"o}wen}},\
  }\href@noop {} {\bibfield  {journal} {\bibinfo  {journal} {Journal of
  Physics: Condensed Matter}\ }\textbf {\bibinfo {volume} {22}},\ \bibinfo
  {pages} {360301} (\bibinfo {year} {2010})}\BibitemShut {NoStop}%
\bibitem [{\citenamefont {Groh}\ and\ \citenamefont {Mulder}(1999)}]{Groh1999}%
  \BibitemOpen
  \bibfield  {author} {\bibinfo {author} {\bibfnamefont {B.}~\bibnamefont
  {Groh}}\ and\ \bibinfo {author} {\bibfnamefont {B.}~\bibnamefont {Mulder}},\
  }\href@noop {} {\bibfield  {journal} {\bibinfo  {journal} {Physical Review
  E}\ }\textbf {\bibinfo {volume} {59}},\ \bibinfo {pages} {5613} (\bibinfo
  {year} {1999})}\BibitemShut {NoStop}%
\bibitem [{\citenamefont {Dong}\ and\ \citenamefont {Evans}(2006)}]{Dong2006}%
  \BibitemOpen
  \bibfield  {author} {\bibinfo {author} {\bibfnamefont {H.}~\bibnamefont
  {Dong}}\ and\ \bibinfo {author} {\bibfnamefont {G.~T.}\ \bibnamefont
  {Evans}},\ }\href {\doibase 10.1063/1.2397076} {\bibfield  {journal}
  {\bibinfo  {journal} {The Journal of Chemical Physics}\ }\textbf {\bibinfo
  {volume} {125}},\ \bibinfo {pages} {204506} (\bibinfo {year}
  {2006})}\BibitemShut {NoStop}%
\bibitem [{\citenamefont {van Teeffelen}\ \emph {et~al.}(2009)\citenamefont
  {van Teeffelen}, \citenamefont {Backofen}, \citenamefont {Voigt},\ and\
  \citenamefont {L\"{o}wen}}]{VanTeeffelen2009}%
  \BibitemOpen
  \bibfield  {author} {\bibinfo {author} {\bibfnamefont {S.}~\bibnamefont {van
  Teeffelen}}, \bibinfo {author} {\bibfnamefont {R.}~\bibnamefont {Backofen}},
  \bibinfo {author} {\bibfnamefont {A.}~\bibnamefont {Voigt}}, \ and\ \bibinfo
  {author} {\bibfnamefont {H.}~\bibnamefont {L\"{o}wen}},\ }\href {\doibase
  10.1103/PhysRevE.79.051404} {\bibfield  {journal} {\bibinfo  {journal}
  {Physical Review E}\ }\textbf {\bibinfo {volume} {79}},\ \bibinfo {pages} {1}
  (\bibinfo {year} {2009})}\BibitemShut {NoStop}%
\end{thebibliography}%

\end{document}